\documentclass[aps,showpacs,graphicx]{revtex4}
\usepackage{amssymb}
\usepackage[dvips]{graphicx}
\usepackage[english]{babel}
\usepackage{indentfirst}
\usepackage{amsxtra}
\usepackage{amsmath}
\usepackage{supertabular}
\usepackage{multirow}
\usepackage [mathcal]{eucal}
\def\beq{\begin{equation}}
\def\eeq{\end{equation}}
\def\beqn{ \begin{eqnarray} }
\def\eeqn{ \end{eqnarray} }
\def\s1s2{{ \boldsymbol{\sigma}(i) \cdot \boldsymbol{\sigma}(j) }}
\def\t1t2{{ \boldsymbol{\tau}(i) \cdot \boldsymbol{\tau}(j)  }}
\newcommand{\rud}{{r_{ij}}}

\newcommand{\eqm}{Q}
\newcommand{\mdm}{\mu}

\newcommand{\betwo}{B(E2)$\uparrow$ }
\newcommand{\bmone}{B(M1)$\uparrow$ }
\newcommand{\mone}{M1 } 
\newcommand{\op}{{\cal B}}
\newcommand{\eop}{{\cal E}}
\newcommand{\mop}{{\cal M}}

\newcommand{\bsigma}{\mbox{\boldmath $\sigma$}}
\newcommand{\btau}{\mbox{\boldmath $\tau$}}
\newcommand{\half}{\frac{1}{2}}
\newcommand{\what}[1]{\widehat #1}

\newcommand{\bpi}{{\bf p}}

\newcommand{\threej}[6]{ \left( \begin{array}{ccc}
                               #1 & #2 & #3 \\
                               #4 & #5 & #6 
                             \end{array}
                        \right) } 
\newcommand{\sixj}[6]{ \left\{ \begin{array}{ccc}
                               #1 & #2 & #3 \\
                                #4 & #5 & #6 
                               \end{array}
                        \right\} } 

\usepackage[usenames]{color}
\usepackage{ulem}

\begin{document}

\noindent
\title{Electric quadrupole and magnetic dipole moments of odd nuclei \\ 
near the magic ones in a self-consistent approach}

\author{G. Co'}
\affiliation{Dipartimento di Matematica e Fisica ``E. De Giorgi'',
 Universit\`a del Salento and,
 INFN Sezione di Lecce, Via Arnesano, I-73100 Lecce, ITALY}
\author{V. De Donno}
\affiliation{Dipartimento di Matematica e Fisica ``E. De Giorgi'',
 Universit\`a del Salento, Via Arnesano, I-73100 Lecce, ITALY}
\author{M. Anguiano, R. N. Bernard,  A. M. Lallena}
\affiliation{Departamento de F\'\i sica At\'omica, Molecular y
  Nuclear, Universidad de Granada, E-18071 Granada, SPAIN}
\date{\today}

\bigskip

\begin{abstract}
We present a model which describes the properties of odd-even nuclei
with one nucleon more, or less, with respect to the magic number.  In
addition to the effects related to the unpaired nucleon, we consider
those produced by the excitation of the closed shell core.  By using a
single particle basis generated with Hartree-Fock calculations, we
describe the polarization of the doubly magic-core with Random Phase
Approximation collective wave functions.  In every step of the
calculation, and for all the nuclei considered, we use the same
finite-range nucleon-nucleon interaction. We apply our model to the
evaluation of electric quadrupole and magnetic dipole moments of
odd-even nuclei around oxygen, calcium, zirconium, tin and lead
isotopes.  Our Random Phase Approximation description of the
polarization of the core improves the agreement with experimental
data with respect to the predictions of the independent particle
model. We compare our results with those obtained in first-order
perturbation theory, with those produced by Hartree-Fock-Bogolioubov
calculations and with those generated within the Landau-Migdal theory
of finite Fermi systems.  The results of our universal,
self-consistent, and parameter free approach have the same quality of
those obtained with phenomenological approaches where the various
terms of the nucleon-nucleon interaction are adapted to reproduce some
specific experimental data.  A critical discussion on the validity of
the model is presented.
\end{abstract}

\bigskip
\bigskip
\bigskip


\maketitle

\section{Introduction}
\label{sec:intro}
The description of the angular momenta of odd-even nuclei is one of
the greatest successes of the nuclear shell-model. The angular
momentum and the parity of the single particle (s.p.) level of the
unpaired nucleon correspond to those of the whole nucleus. This
extreme shell model picture, where the s.p. properties are imposed to
the whole interacting many-body system, is weakened when observables
which are not quantized are investigated.  For example, the values of
electric quadrupole and magnetic dipole moments are rather different
from those predicted by the extreme shell model. For these quantities,
the interaction between the unpaired nucleon and the other nucleons
plays an important role, strongly modifying
the pure shell model predictions.

One expects that in odd-even nuclei around doubly-magic ones, the
effects related to the extreme shell model picture, which we shall
name henceforth Independent Particle Model (IPM), are dominant with
respect to those induced by the two-body part of the nuclear
hamiltonian, usually called residual interaction.  This has induced
the development of various perturbative models \cite{goo72,ell77}.  A
class of these models is based on the straightforward application of the
traditional perturbation theory, which, however, has convergence
problems, at least for the terms beyond those of the first order
\cite{li13}.

Another perturbative approach consists in considering the residual
interaction within the framework of the Random Phase Approximation
(RPA) theory.  This approach is based on the extension of the Landau
theory of quantum liquids to finite Fermi systems (FFS) done by Migdal
\cite{mig67}. This theory has been applied with great success to
describe the properties of odd-even nuclei in the region of $^{208}$Pb
\cite{rin73,bau73,spe77}. In these works the s.p. wave functions are
generated by diagonalizing a Woods-Saxon well, and the residual
interaction is a zero-range Landau-Migdal force \cite{mig67}.  More
recently, this approach has been extended to use the same interaction
in the production of the s.p. wave functions and in the RPA
calculations \cite{bor08,bor10,tol11,tol12,voi12}.  Furthermore, also
pairing effects have been considered. This self-consistent theory of
FFS has been applied also to the tin and to the calcium isotopes. In
these approaches, the use of the concept of quasi-particles implies
the definition of the effective charge which contains free parameters
whose values are chosen to have a good description of the data.  In
the phenomenological approach of Refs. \cite{rin73,bau73,spe77} also
the constants defining the strength of the various channels of the
Landau-Migdal nucleon-nucleon effective interaction have been used as
free parameters. In this case, the values of these parameters are
related to the dimension of the s.p. configuration space. The
application of these methods to a different region of the nuclear
chart, or, more simply, to a different s.p. configuration space,
requires a new selection of the free parameter values.  The
self-consistent FFS approach of
Refs. \cite{bor08,bor10,tol11,tol12,voi12} does not have cut-off
problems in the RPA calculations, since it considers the full
s.p. configuration space, continuum included. The problems arise in
the pairing calculations, where the free parameter is the value of the
cutoff energy used as regulator, in analogy to the renormalisation
procedure adopted in effective field theories.

Inspired by these works, we propose here an extension of the RPA
formalism to construct a fully self-consistent approach.  For each
nucleus considered, we use a set of s.p. wave functions generated by a
Hartree-Fock (HF) calculation. The same, effective, nucleon-nucleon
interaction used in HF is adopted in the RPA calculations, which
allows the evaluation of the odd-even nucleus observables.  The
parameters of the force, which we consider having finite-range, have
been chosen once forever in a global HF fit of properties mainly
related to the ground state of a large set of nuclei in all the
regions of the nuclear chart \cite{cha07t}. The use of a finite-range
interaction ensures the stability of our results whose values, after
convergence has been reached, do not depend any more on the size of
the s.p. configuration space.

The basic ideas of our work are relatively
simple and straightforward.  First, we use effective nucleon-nucleon
interactions whose parameters have been chosen to reproduce data
which are not directly related to the observables we
investigate. Second, the chain of calculations needed to arrive at the
final results does not contain any additional free
parameter, but they depend exclusively on the effective
nucleon-nucleon interaction. To be sure of
  identifying features related to the theory, and not to the specific
choice of the interaction or to the nucleus investigated, we carried
out calculations with different 
nucleon-nucleon forces, for a set of nuclei in
a large region of the nuclear chart. We
mainly concentrate our investigation on two observables, the electric
quadrupole, $\eqm$, and the magnetic dipole, $\mdm$, moments of a
selected set of odd-even nuclei.  The evaluation of these quantities
requires, respectively, the description of the $2^+$ and $1^+$
excitations of the even-even core nuclei which we obtain by using a
RPA approach \cite{don14a}.

In Sec.~\ref{sec:model} we present the basic ideas our model. In this
section we also specify the form of the electromagnetic operators that
we have adopted to describe $\eqm$ and $\mdm$. The expression of the
operator for $\eqm$ is simple and straightforward. More complex is the
situation for $\mdm$, where, in addition to the traditional one-body
operator, we consider also two-body currents generated by the exchange
of charged pions. We shall refer to these latter currents
as meson exchange currents (MEC).
In Sec.~\ref{sec:sa} we discuss some details of the calculations
related to the specific applications of the model.  The results
obtained for $\eqm$ and $\mdm$ are presented in Sect.~\ref{sec:res}.
In both cases we also study the excitation of the even-even core
nuclei for the two multipolarities involved.  In Sec.~\ref{sec:conc}
we summarize the main results of our study, and
we draw our conclusions.

\section{The model}
\label{sec:model}
The starting point of our model is the construction of the basis of
s.p. states $|\phi_\alpha \rangle$, which we generate by solving the
HF equations.  We used the symbol $\alpha$ to indicate all the quantum
numbers characterizing the s.p. state, i.e. the principal
quantum number $n_\alpha$, the orbital angular momentum
$l_\alpha$, the total angular momentum $j_\alpha$, its third
component $m_\alpha$, and the third component of the isospin
$t_\alpha=-1/2$ for protons, and $t_\alpha=1/2$ for neutrons.

In the IPM picture, all the s.p. states below the Fermi energy are
completely occupied, and those above it are empty. We indicate with
$|\Phi_0 \rangle$ the Slater determinant describing the IPM ground
state of the doubly magic nucleus composed by $A$ nucleons.  By
definition, the RPA ground state of the doubly magic nucleus, which we
indicate as $|\Psi_0 \rangle$, contains correlations beyond the IPM.

The odd-even nuclei we want to describe are obtained by adding, or
subtracting, one nucleon to the doubly-magic $A$-nucleon system.  We
label the states of these nuclei as $|A \pm 1; \alpha \rangle$, where
the symbol $\alpha$ indicates the set of quantum numbers
characterizing the s.p. state of the added, or subtracted, nucleon.

We describe the response of the odd-even nucleus to the perturbation
induced by an external operator $\op$ by considering separately two
effects.  The first one is the action of the external operator on the
unpaired nucleon, while the doubly-magic core remains
unperturbed. The second effect considers the interaction of the
external probe with one of the nucleons of the doubly magic core.  In
this case, the whole nucleus responds to the external perturbation
because the nucleon struck by the external probe interacts strongly
with all the other nucleons, included the unpaired one. 

We express the expectation value of the external operator $\op$ between
two states of the odd-even nucleus with one nucleon more than the 
doubly-magic one as \cite{mig67,rin73,bau73,spe77}
\beq
\langle  \,A + 1; \alpha\, |\, \op \,|\,A + 1; \beta \,\rangle \, = \,
\langle  \, \Phi_0 \,|\, a_\alpha \, \op \,  a^+_\beta \,|\, \Phi_0 \,\rangle  \,
+ \,\langle \, \Psi_0 \, |\, a_\alpha \,a^+_\beta \,V_{\rm res} \, {\cal P}_A(\epsilon_{\alpha\beta})\,
\op \, |\, \Psi_0 \,\rangle
\, ,
\label{eq:start}
\eeq
where $a^+$ and $a$ are the usual creation and annihilation operators,
$V_{\rm res}$ is the residual nucleon-nucleon interaction, and ${\cal
  P}_A$ is a propagator which describes the excitation of the doubly
magic core and depends on
$\epsilon_{\alpha\beta}=\epsilon_\alpha-\epsilon_\beta$, the
difference between the energies of the two states of the $A+1$ system.
A similar equation can be written for the $A-1$ nucleus by exchanging
creation and annihilation operators.

In our approach, the excitation of the doubly magic core is described
by using the RPA theory.  In terms of an excitation operator
$Q^+_\nu$, we define the excited states of the $A$-nucleon system as
\beq
|\,\Psi_\nu \, \rangle \, = \, Q^+_\nu \, |\, \Psi_0 \, \rangle 
\, ,
\label{eq:estate}
\eeq
while the relation 
\beq
Q_\nu \, |\,\Psi_0 \,\rangle \, = \, 0 
\label{eq:gstate}
\eeq
defines the ground state. The label $\nu$ indicates 
all the quantum numbers needed to characterize the excited state. 
The RPA excitation operator 
is defined as 
\beq
Q^+_\nu \, =\, \sum_{ph} \,
\left( \,X^\nu_{ph} \,a^+_p \,a_h \,-\,  Y^\nu_{ph} \,a^+_h \,a_p\,
\right)
\, ,
\label{eq:q}
\eeq
where the index $p$ refers to s.p. states above the Fermi surface
(particles) and $h$ to s.p. states below it (holes). 
The RPA propagator can be written as \cite{mig67,spe77,fet71}
\beq
{\cal P}^{\rm RPA}_A (\epsilon_{\alpha \beta}) \,= \, \sum_{\nu}\,
Q^+_\nu \, |\,\Psi_0 \,\rangle \,
\left(\, \frac{1}{\epsilon_{\alpha \beta} \, -\,  \omega_\nu} \,
      - \, \frac{1}{\epsilon_{\alpha \beta} \,+ \, \omega_\nu} \, \right)
\langle \, \Psi_0 \,| \, Q_\nu \, ,
\eeq
where $\omega_\nu$ is the excitation energy of the state 
$|\,\Psi_\nu \, \rangle$. By using this propagator,  we express 
the second term in the r.h.s. of Eq. (\ref{eq:start}) as
\beqn
\nonumber
\langle  \, \Psi_0 \,| \,a_\alpha \,a^+_\beta \, 
V_{\rm res} \, {\cal P}^{\rm RPA}_A(\epsilon_{\alpha \beta}) \,
\op \,|\,\Psi_0 \,\rangle &=& \\ 
&& \hspace*{-3.cm} =\,
 \langle \, \Psi_0\,|\, a_\alpha\, a^+_\beta\, V_{\rm res}\, \sum_{\nu}
\, Q^+_\nu \, |\, \Psi_0 \, \rangle \, 
\left(\, \frac{1}{\epsilon_{\alpha \beta}\, -\, \omega_\nu} \,
      - \, \frac{1}{\epsilon_{\alpha \beta} \,+\, \omega_\nu}\, \right)\, 
\langle \,\Psi_0\, |\, Q_\nu \, \op\, |\, \Psi_0 \, \rangle
\\ 
& & \hspace*{-3.cm} =\, 
\sum_{\nu} \, \langle \, \Psi_0\,|\, \left[ \, a_\alpha\, a^+_\beta\, V_{\rm res}\, , Q^+_\nu \, \right] \, |\, \Psi_0 \, \rangle 
\nonumber  \left(\, \frac{1}{\epsilon_{\alpha \beta}\, -\, \omega_\nu} \,
      - \, \frac{1}{\epsilon_{\alpha \beta} \,+\, \omega_\nu}\, \right)\, 
\langle \,\Psi_0\, |\, \left[ \,Q_\nu \, , \op \, \right] \, |\, \Psi_0 \, \rangle
\, ,
\eeqn
where we have exploited Eq. (\ref{eq:gstate}) to insert the commutator
$[\,,]$ between operators, with the 
aim of using the quasi-boson-approximation \cite{rin80}:
\beqn
\langle  \, \Psi_0 \,| \,a_\alpha \,a^+_\beta \, 
V_{\rm res} \, {\cal P}^{\rm RPA}_A(\epsilon_{\alpha \beta}) \,
\op \,|\,\Psi_0 \,\rangle &\simeq& \\\nonumber 
&& \hspace*{-3.cm} \simeq \,
 \sum_{\nu} \, \langle \, \Phi_0\,|\, \left[ \, a_\alpha\, a^+_\beta\, V_{\rm res}\, ,Q^+_\nu \, \right] \, |\, \Phi_0 \, \rangle 
\left(\, \frac{1}{\epsilon_{\alpha \beta}\, -\, \omega_\nu} \,
      - \, \frac{1}{\epsilon_{\alpha \beta} \,+\, \omega_\nu}\, \right)\, 
\langle \,\Phi_0\, |\, \left[ \,Q_\nu \, , \op \, \right]\, |\, \Phi_0 \, \rangle
\, .
\eeqn

The operator describing the action of the external probe on the
nuclear system can be expressed in terms of a multipole
expansion. Each term of this expansion, ${\cal B}_{JM}$, is
characterized by the angular momentum $J$, and its projection $M$ on
the quantization axis. In our calculations we consider the spherical
symmetry of the problem.  We use a set of s.p. wave functions with the
following angular momentum coupling structure:
\beq
\label{eq:spwf}
\phi_\alpha({\bf r}) \,=\, R^{t_\alpha}_{n_\alpha l_\alpha j_\alpha}(r)\,
\sum_{\mu_\alpha s_\alpha} \, \langle \, l_\alpha \, \mu_\alpha \, \half \, s_\alpha \, | \, j_\alpha \, m_\alpha \, \rangle \,
Y_{l_\alpha \mu_\alpha}(\theta,\phi) \, \chi_{s_\alpha}
\, ,
\eeq
where $(r,\theta,\phi)$ are the usual polar coordinates, $Y_{l \mu}$
are the spherical harmonics, $\chi$ the Pauli spinor, and the symbol
$\langle|\rangle$ indicates the Clebsch-Gordan coefficient. We
calculate the matrix elements by applying the Wigner-Eckart theorem
with the phase conventions of Ref. \cite{edm57}, and we sum on all the
possible values of $m_\alpha$ and $m_\beta$. For the matrix element in
Eq. (\ref{eq:start}) we obtain
\beqn
\nonumber
\langle \,A+1; \alpha \, ||\, \op_J \, ||\, A+1; \beta \, \rangle &=& \\ 
&& \hspace*{-2.cm} =\,
\langle \, \phi_{\alpha} \, ||\, \op_J \, ||\, \phi_{\beta}\, \rangle \,  
+ \, \delta_{\alpha,\beta} \, \sum_{h} \,
\langle \, \phi_{h} \, || \, \op_J \, ||\, \phi_{h} \, \rangle 
\label{eq:main}
\\
\nonumber
&&\hspace*{-1.5cm} +\,
\sum_{\nu}  \, 
\Big[\, {\cal D}^{\nu}_{\alpha \beta}\, \sum_{p h} \, {\cal A}_{ph}^\nu \, 
\langle \, \phi_{p}\, || \, \op_J \, ||\,\phi_{h} \, \rangle 
\,+\,
{\cal G}^{\nu}_{\alpha \beta}\, \sum_{p h} \, (-1)^{j_{h}+j_{p}} \, {\cal A}_{ph}^\nu \,  
\langle \, \phi_{p}\, || \, \op_J \, ||\,\phi_{h} \, \rangle  \Big] 
\, ,
\eeqn
where the two kernels are defined as
\beq
{\cal D}^{\nu}_{\alpha \beta} \,=\, \sum_{ph}
v^J_{\alpha \beta p h} \left(
\frac  { X^\nu_{p h}} 
{\epsilon_{\alpha \beta} - \omega_\nu} +
(-1)^{j_{h}+j_{p}}
\frac  {Y^\nu_{p h}}
{\epsilon_{\alpha \beta} + \omega_\nu} \right) 
\label{eq:kern1}
\eeq
and
\beq
{\cal G}^{\nu}_{\alpha \beta} \,=\, \sum_{p h}
v^J_{\alpha \beta h p} \left(
\frac  { X^\nu_{p h}} 
{\epsilon_{\alpha \beta} + \omega_\nu} +
(-1)^{j_{h}+j_{p}}
\frac  {Y^\nu_{p h}}
{\epsilon_{\alpha \beta} - \omega_\nu} \right) \, ,
\label{eq:kern2}
\eeq
with
\beq
{\cal A}_{ph}^\nu \, = \, X^\nu_{p h} \,+ \,(-1)^{j_{h}-j_{p}} \,Y^\nu_{p h} \, .
\eeq
In Eq. (\ref{eq:main}) the double bars indicate that in the matrix
elements between the two s.p. wave functions the angular part is
evaluated in terms of reduced matrix elements. The 
calculation of the radial integrals is understood.

The expression of the interaction terms $v^J$ is analogous to that 
of the RPA \cite{fet71}
\beqn
\nonumber
v^J_{\alpha \beta \gamma \delta} & = & \sum_K \, (-1)^{K+j_\beta+j_{\delta}} \, \what{K} \,
\sixj {j_\alpha}{j_\beta}{J}{j_{\delta}}{j_{\gamma}}{K} \\
& & \times \, \Big[
\langle  \, j_\alpha\, j_{\gamma}\, K \, \|\, V_{\rm res}\, \|\, j_\beta\, j_{\delta} \, K \, \rangle \, 
- \, (-1)^{j_\beta+j_{\delta}-K}\,  
\langle \, j_{\delta} \, j_\beta \,  K\, \|\, V_{\rm res}\, \|\,  j_\alpha \, j_{\gamma} \,  K \, \rangle
\Big]
\, .
\label{eq:vj}
\eeqn
In the above equation we used the Wigner 6-$j$ symbol \cite{edm57},
and $\what{j} \equiv \sqrt{2j+1}$, where $j$ is referred to an angular
momentum.
The interaction $V_{\rm res}$ is described as
\beqn
V_{\rm res} &=& v_1(\rud) \, +\, v_2(\rud)\,\t1t2 \,
  + \, v_3(\rud)\,\s1s2 \, + \, 
v_4(\rud)\,\s1s2 \,\t1t2  \\ 
\nonumber
&& +\,  t_\rho \left[ 1 - \t1t2 \right] \,
\rho^{1/3} \, \left(\frac{r_i + r_j}{2} \right)
\delta(\rud) 
\, + \, 2\, i\, W_0 \,
\left[ \overleftarrow{\bpi}_{ij} \times \delta(r_{ij}) 	\,
  \overrightarrow{\bpi}_{ij} \right] 
\cdot \mathbf{S} + v_{\rm Coul}(\rud)
\nonumber
\, ,
\label{eq:ginteraction}
\eeqn
where $t_\rho$ and $W_0$ are real constants, $v(\rud)$ are scalar
functions of the distance between the two nucleons, and 
$v_{\rm Coul}(\rud)$ is the Coulomb interaction acting between two protons.
In the above equations $\bsigma$ and $\btau$ are the usual Pauli
matrices for the spin and isospin operators respectively, $\rho$ is
the nucleon density, $\mathbf{S} = [\bsigma(i) + \bsigma(j)] / 2$ and
${\bf p}$, is the relative momentum of the interacting pair with the
arrows indicating the side on which the operator acts.  It is worth
pointing out that the same effective nucleon-nucleon interaction is
used in the HF and in all the RPA calculations.

The sum over $\nu$ in Eq. (\ref{eq:main}) runs over all the excited
states of the $A$-nucleon core with the same multipolarity and parity
of the transition operator.  Eq. (\ref{eq:main}) is the basic expression
of our model and is analogous to those obtained in Refs.
\cite{rin73,bau73,spe77} within the FFS theory.

The results obtained by using the whole expression in
Eq. (\ref{eq:main}) have been labelled as RPA. We compare them with
those of the IPM, i.e. those provided by the first
term of Eq. (\ref{eq:main}). In addition we consider also the
  results obtained within the framework of 
the traditional time-independent first order
perturbation theory (FOPT), which we obtain by calculating 
\beqn
\nonumber
\langle \,A+1; \alpha \, ||\, \op_J \, ||\, A+1; \beta \, \rangle_{\rm
  FOPT}  &=& 
\langle \, \phi_{\alpha} \, ||\, \op_J \, ||\, \phi_{\beta}\, \rangle \,  
+ \, \delta_{\alpha,\beta} \, \sum_{h} \,
\langle \, \phi_{h} \, || \, \op_J \, ||\, \phi_{h} \, \rangle 
\\
&&
+\,
\sum_{p h}  \, v^J_{\alpha \beta h p} \,
\left(
\, \frac{1}{\epsilon_{\alpha \beta} - \epsilon_{ph}} \, 
+ \, \frac{1}{\epsilon_{\alpha \beta} + \epsilon_{ph}} \, 
\right) \,
\langle \, \phi_p \, ||\, \op_J \, ||\, \phi_h \, \rangle 
\, .
\label{eq:FOPT}
\eeqn

For nuclei with one nucleon less than the doubly magic ones we obtain
equations similar to \eqref{eq:main} and \eqref{eq:FOPT} with an
additional overall phase. 

The first operator we consider in this paper is a one-body
electric operator,
\beq
\eop_{JM} \, = e \sum_i^A
\,\left[ \frac{1}{2} - t_i \right]\,r^J_i \, Y_{JM}(\Omega_i) 
\, ,
\eeq
where $e$ is the elementary charge. We express
the reduced s.p. matrix element
of this operator as
\beqn
\langle  \,\phi_{\alpha}\, \| \, \eop_J \, \|\, \phi_{\beta} \, \rangle
&=& \frac{e}{\sqrt{4\pi}} \,\left( \frac{1}{2} - t_\alpha \right) \, \delta_{t_\alpha
  t_\beta} \,  
(-1)^{j_\alpha -1/2} \, \xi(l_\alpha + l_\beta +J) \,  \what{j_\alpha} \, \what{j_\beta} \, \what{J} \,
\threej{j_\alpha} {J} {j_\beta} {-\half} {0} {\half}
{\cal I}^J_{\alpha \beta} 
\, .
\label{eq:spel}
\eeqn
In the above expression we used the Wigner 3-$j$ symbol \cite{edm57}, and
$\xi(l)=1$ if $l$ is even and zero otherwise.
We have indicated with ${\cal I}_{\alpha \beta}^J$ the integral 
\beq
{\cal I}^J_{\alpha \beta} \, =\,  \int {\rm d}r \, r^{J+2} \, R^{t_\alpha}_{n_\alpha l_\alpha j_\alpha}(r)\, R^{t_\beta}_{n_\beta l_\beta j_\beta}(r)
\, .
\eeq
Specifically, we calculate the electric quadrupole moment
 $\eqm$ defined as 
\beq
\eqm = \sqrt{\frac{16 \pi}{5}} 
\threej{j_\alpha} {2} {j_\alpha} {-j_\alpha} {0} {j_\alpha}
\langle  \,A \pm 1; \alpha\, \| \, \eop_{2} \, \| \,A \pm 1; \alpha \,\rangle 
\, .
\label{eq:quadrupole}
\eeq

The second operator we consider in this work is the magnetic
operator, composed by one- and two-body terms. 
We express the one-body (OB) part of the magnetic operator as
\beq
\mop^{\rm OB}_{JM} \, =\, 
\mu_N \, \sum_{i=1}^A 
\left[ \frac{2}{J+1}\, g_l(t_i) \, {\bf l}(i) \,+\, 
g_s(t_i) \,\bsigma(i) \right]
\cdot \nabla_i \left[ r^J_i \, Y_{JM}(\Omega_i) \right]
\, ,
\label{eq:mob}
\eeq where $ \mu_N $ is the nuclear magneton, and we have for the 
gyroscopic constants
the values $g_l$=1 and 0, and 
$g_s$ = 2.793 and $-1.913$, for protons and neutrons, respectively.
The expression for the s.p. matrix element is
\beqn
\nonumber
\langle  \,\phi_{\alpha}\, \| \,\mop^{\rm OB}_J \, \|\, \phi_{\beta} \, \rangle
&=& \frac{1}{\sqrt{4\pi}} \, \delta_{t_\alpha t_\beta}\, 
(-1)^{j_\beta-1/2+J} \, 
\xi(l_\alpha + l_\beta +J+1) \, 
 \what{j_\alpha} \, \what{j_\beta} \, \what{J} \, \threej{j_\alpha} {j_\beta} {J} {\half} {-\half} {0} \\
&& \times \,
(J - \kappa_{\alpha \beta}) \,
\left[g_l(t_\alpha) \left( 1+ \frac {\kappa_{\alpha \beta}}{J+1} \right) \,  -\, g_s(t_\alpha)\right]
{\cal I}^{J-1}_{\alpha \beta} 
\, ,
\label{eq:spmag}
\eeqn
where we have defined
\beq
\kappa_{\alpha \beta} \,=\, (l_\alpha - j_\alpha)\, (2j_\alpha+1)\,
+\, (l_\beta - j_\beta) \, (2j_\beta+1)
\label{eq:kappa}
\, .
\eeq

%
\begin{figure}[ht]
\begin{center}
\includegraphics[scale=0.6, angle=0] {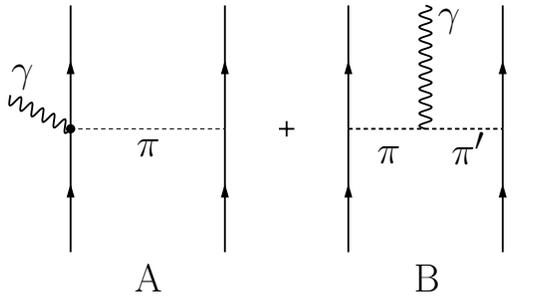} 
\vspace*{-0.3cm}
\caption{Schematic representation of the MEC diagrams 
considered in this work. The wavy lines represent the external electromagnetic probe, 
the dashed lines the exchanged charged pion, and the
full lines the nucleons. The A diagram is normally called 
seagull and the B diagram pionic.
} 
\label{fig:MEC}
\end{center}
\end{figure}

In addition to the one-body operator, we consider two-body terms
generated by the exchange of a single, charged pion, the so-called
seagull, $\mop^{\rm s}$, represented by the A diagram of
Fig. \ref{fig:MEC}, and pionic, $\mop^{\pi}$, represented by the B
diagram of the same figure. We give in Appendix \ref{sec:mec} the
expressions of the matrix elements of these two MEC operators.
In the FFS theory, these MEC corrections to the OB operator 
are taken into account by considering the effective charge of the 
quasi-particle and by including an effective tensor term.

Specifically, we calculate the magnetic dipole moment 
defined as
\beq 
\mu \, =\, \sqrt{\frac{4 \pi}{3}} \threej{j_\alpha}
     {1} {j_\alpha} {-j_\alpha} {0} {j_\alpha} \langle \,A \pm 1;
     \alpha\, \| \, \mop_{1} \, \| \,A \pm 1; \alpha \,\rangle \
\, ,
\label{eq:dipole}
\eeq
where $\mop_1 = \mop_1^{\rm OB} + \mop_1^{\rm MEC}$. 
In IPM calculations, by
considering $\mop_1 = \mop_1^{\rm OB}$, we obtain the well known Schmidt
values \cite{sch37,rin80} which depend only on the 
angular momentum $j$ of the s.p. level of the unpaired nucleon and are given by:
\begin{equation}
\label{eq:schmidt}
\mu_{\rm Sch} \,  =\, \left\{
\begin{array}{ll}
\displaystyle \left[ \left(j-\frac{1}{2} \right) \, g_l \,+\, g_s \right]\, \mu_N \, , &\mbox{for} \,\, j=l+\frac{1}{2} \, , \\[3mm]
\displaystyle \frac{j}{j+1} \left[ \left(j+\frac{3}{2} \right) \, g_l \,-\, g_s \right]\, \mu_N \, , &\mbox{for} \,\, j=l-\frac{1}{2} \, .
\end{array}
\right.
\end{equation}

A remarkable difference between our calculations and those based on
the FFS theory is that we use bare, not effective, electromagnetic
operators, therefore we do not include free parameters in our theory.
Furthermore, we do not include in the magnetic operator \eqref{eq:mob}
the tensor terms often added to take care of magnetic dipole $\Delta l
= 2$ transitions in the IPM picture \cite{boh69,bau73,bor08}.

%
%

\section{Specific applications}
\label{sec:sa}
As pointed out in the Introduction, we carried out our calculations by
using two different parameterizations of the finite-range density
dependent Gogny interaction.  These define the scalar functions
$v(\rud)$ of Eq. \eqref{eq:ginteraction}, and the values of the
$t_\rho$ and $W_0$ constants.  We have chosen the traditional D1S
force \cite{ber91} and the more recent D1M one \cite{gor09}, which
improves the behavior of the neutron matter equation of state at high
density values.  We would like to emphasize the fact that the
parameters of these interactions have been chosen once forever in a
fit process, described in detail in Ref. \cite{cha07t}, and they have
not been modified in the present work. Since the effective
nucleon-nucleon interaction is the only input of our calculations, we
may state that they are parameter free. 

For each nucleus considered, the calculations of $\eqm$ and $\mdm$ are
carried out in three steps. In the first one we generate the set of
s.p. wave functions by means of a HF calculation.  In the second step,
for a selected multipole excitation of angular momentum $J$ and parity
$\Pi$, we solve the RPA equations. In the third step we calculate the
expectation value of the operators $\eop_2$ and $\mop_1$ by using the
expression (\ref{eq:main}) or the ones for IPM and FOPT.

From the computational point of view, it is necessary to ensure the
numerical convergence of each of these three steps.  In our
calculations we use a set of s.p. wave functions with bound boundary
conditions at the edge of an $r$-space box. In this manner all the
s.p. states are bound, even those with positive energy. In the common
jargon these are called discrete calculations.  For the first two
steps, the numerical stability of the results is related to the
dimensions of the $r$-space integration box and to the size of the
s.p. configuration space used in the discrete RPA calculations. We
handle these problems by using the strategy described in
Ref. \cite{don14b}, i.e. by choosing the size of the integration
boxes and the dimensions of the s.p. configuration spaces such as the
centroid energies of the electric giant dipole resonances do not
change by more than 0.5 MeV if the value of any of the two parameters
is increased. In this respect, the most demanding calculations are
those carried out for the $^{208}$Pb nucleus where we have used a box
radius of 25 fm and a maximum s.p. energy of 100 MeV.

\begin{figure}[th]
\begin{center}
\includegraphics[scale=0.4, angle=0] {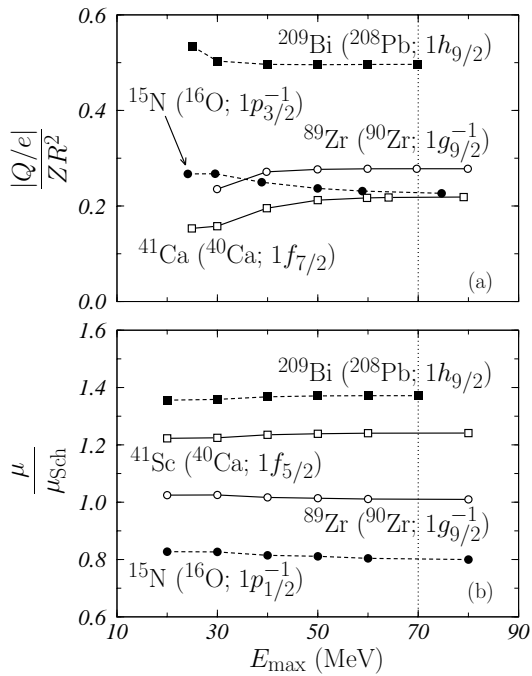} 
\vspace*{-0.3cm}
\caption{\small Convergence test of the stability of our results against 
$E_{\rm max}$, the upper limits of the sum on $\nu$ in Eq. (\ref{eq:main}),
 for $\eqm$, panel (a), and $\mu$, panel (b), 
as given by Eqs. (\ref{eq:quadrupole}) and (\ref{eq:dipole}), respectively.
The vertical lines indicate the 70 MeV values which we have adopted as
the value of $E_{\rm max}$ in our calculations.} 
\label{fig:conv}
\end{center}
\end{figure}

The stability of the results of the third step is related to the
maximum value of the RPA excitation energy, $E_{\rm max}$, which is
the upper limit of the sum on $\nu$ in Eq. (\ref{eq:main})
and of the sum on $ph$ in Eq. \eqref{eq:FOPT}.  We
carried out convergence tests for each nucleus considered.  An example
of these tests is presented in Fig. \ref{fig:conv}, where we show the
values of the quantities $|\eqm/e| / Z R^2$, with $R=1.2\,A^{1/3}$, in
the upper panel, and $\mdm / \mdm_{\rm Sch}$, in the lower one, for
the ground states of various nuclei, as a function of $E_{\rm max}$.
We observe that in Fig. \ref{fig:conv} all the results have already
converged at $E_{\rm max}$ = 70 MeV, and this is the minimum value of
$E_{\rm max}$ we adopted in our calculations.

%

\section{Results}
\label{sec:res}

We have applied the model presented in the previous sections to
various nuclei in different regions of the nuclear chart. The core
nuclei have been selected to have closed shells and spherical
symmetry, therefore, in these nuclei, deformation and pairing effects
do not play any role \cite{del10,ang14}.  We have identified isotopes
of oxygen, calcium, zirconium, tin and lead endowed with these
characteristics.  We have investigated the role of the pairing in
these nuclei by carrying out calculations in a HF plus Bardeen Cooper
and Schrieffer framework \cite{ang14}. The protons and neutrons energy
gaps between particle and hole states are so large that we did not
find pairing effects. We found only one exception, the case of the
protons in $^{90}$Zr, where the particle fluctuation number is not
exactly zero. The evaluation of the occupation probabilities of the
single particle levels indicates deviations of few percent from that
predicted by a sharp Fermi distribution. Only the proton $2p_{1/2}$
state, which in IPM does not contributes to the $\eqm$, is sensitively
affected by the pairing with an occupation probability of 85\%. This
s.p. state does not contributes to the $\eqm$.

We first present the results obtained for $\eqm$, and, in a second
step, those related to $\mdm$.

\subsection{The $2^+$ excitations and the $\eqm$ values.}

We show in Table \ref{tab:E2} the excitation energies and the \betwo
values of the first $2^+$ state obtained in our RPA calculations for
the various nuclei studied, and we compare them with the available
experimental values taken from the compilations of
Refs. \cite{led78,bnlw,ram01}.  We notice a large difference between
the RPA values and the experimental ones in $^{16}$O and $^{40}$Ca
nuclei.  The experimental energies are smaller than those predicted
by our calculations, and the experimental \betwo values much larger,
especially in $^{40}$Ca.  Evidently, for $^{16}$O and $^{40}$Ca, the
lowest $2^+$ states cannot be properly described by the RPA approach
since their structure is more complex than a simple combination of
one-particle one-hole excitations. In effect, the experimental
excitation spectrum of $^{16}$O presents at least other six 2$^+$
states up to 15 MeV, and that of $^{40}$Ca about fifty of them below
10 MeV \cite{led78,bnlw}.

The situation for the other nuclei is different. We found reasonable
agreement with the available empirical data in $^{48}$Ca, $^{132}$Sn
and $^{208}$Pb.  The 2$^+$ states in these nuclei are dominated by
particle-hole (ph) transitions between bound s.p. states.  The
simplest case is that of the $^{48}$Ca nucleus where we found that the
neutron ($2p_{3/2} \, 1f_{7/2}^{-1}$) transition is the 
dominant component of the first $2^+$ excited state.  We identify
analogous situations in $^{132}$Sn, with the proton ($2d_{5/2} \,
1g_{9/2}^{-1}$) and neutron ($2f_{7/2} \, 1h_{11/2}^{-1}$)
transitions, and in $^{208}$Pb with the proton ($2f_{7/2} \,
1h_{11/2}^{-1}$) and ($1h_{9/2}\, 1h_{11/2}^{-1}$) and neutron
($2g_{9/2} \, 1i_{13/2}^{-1}$) transitions.  In these nuclei, the
$2^+$ strength is concentrated in the s.p.  transitions above
mentioned with little fragmentation produced by the coupling with the
continuum and/or with more complex excitations modes.

%
%
\begin{table}[t]
\begin{center}
{\footnotesize
\begin{tabular}{ccccccccc}
\hline \hline 
&~~~~~~~~~~&  \multicolumn{3}{c}{$\omega$ (MeV)} &~~~~~~~~~~~~~& \multicolumn{3}{c}{\betwo  (e$^2\,$fm$^4$)} \\ \cline{3-5} \cline{7-9}
nucleus && D1M &  D1S &  exp && D1M &  D1S &  exp \\
\hline 
  $^{16}$O && 17.61  & 18.13  &   6.92 && 16.76  & 22.85  &  40.64 \\
  $^{22}$O &&  2.34  &  2.65  &    3.19  && 15.84  & 14.72 &  28.12 \\
  $^{24}$O &&  4.10  &  3.97  &             &&  4.95  &  4.50  &    \\
  $^{40}$Ca&& 13.41  & 14.63 &  3.90  &&  3.08  &  3.66 & 99.17 \\
  $^{48}$Ca&&  3.80  &  4.04 &  3.83   && 91.65  & 87.91 & 95.32 \\
  $^{60}$Ca&&  5.82  &  5.63 &           && 0.39   &  0.40  &   \\
  $^{90}$Zr&& 4.67   & 4.95   &   2.19 &&353.5  &350.4 &  610.4 \\
 $^{100}$Sn&& 4.13   & 4.40 &          &&1950    &2055  &   \\
 $^{132}$Sn&& 4.26   & 5.08 &  4.04   && 1574   & 1444   &   \\
 $^{208}$Pb&& 4.91 & 5.05    &  4.09  && 4083   & 4244   &  3003 \\
\hline \hline
\end {tabular}
}
\end{center}
\vspace*{-0.5cm}
\caption{\small Excitation energies, $\omega$ and \betwo
  values of the lowest 2$^+$ state for the core nuclei
  considered. The results obtained with the two interactions we have
  adopted are shown. 
 The experimental values are taken from Refs. \cite{led78,bnlw,ram01}.   
}
\label{tab:E2}
\end{table}

\begin{table}[!hb]
\begin{center}
{\scriptsize
\begin{tabular}{ r  llrr r  llrr }
\hline \hline 
core & nucleus  &s.p. state & D1M  & D1S & core & nucleus & s.p. state & D1M  & D1S  \\ 
\hline 
 $^{   16}$O  &  $^{  15}$N &   $1p_{ 3/2}^{-1}$ &     2.07 &         2.17 &  
 $^{  100}$Sn &  $^{ 99}$In &   $1g_{ 9/2}^{-1}$ &    22.48 &        23.03  \\
 $^{    }$ &  $^{  17}$F    &   $1d_{ 5/2}$     &    -4.50 &        -4.74 &                            
 $^{    }$ &  $^{ 101}$Sb   &   $1g_{ 7/2}$     &   -25.04 &       -25.32      \\                     
 $^{    }$ &  $^{  17}$O    &   $1d_{ 5/2}$     &    -1.61 &        -1.55 &                           
 $^{    }$ &  $^{ 99}$Sn    &   $1g_{ 9/2}^{-1}$ &    16.21 &        16.31     \\   \cline{1-5} 
 $^{  22}$O &  $^{  21}$N   &   $1p_{ 3/2}^{-1}$ &     3.69 &        3.55  &                  
 $^{    }$ &  $^{ 101}$Sn   &   $1g_{ 7/2}$     &   -19.12 &       -18.95      \\   \cline{6-10} 
 $^{    }$ &  $^{  23}$F    &   $1d_{ 5/2}$     &    -6.53 & -6.35 & 
 $^{ 132}$Sn &  $^{ 131}$In &   $1g_{ 9/2}^{-1}$ &    18.61 &     17.70  \\                     
 $^{  24}$O & $^{  23}$N    &   $1p_{ 3/2}^{-1}$ &     2.34 &         2.47 & 
 $^{    }$ &  $^{ 133}$Sb   &   $1g_{ 7/2}$     &   -20.13 &    -23.02\\  
 $^{    }$ &  $^{  25}$F    &   $1d_{ 5/2}$     &    -4.58 &        -4.82 &            
 $^{    }$ & $^{ 131}$Sn    &   $1h_{11/2}^{-1}$ &    14.65 &     12.41 \\  \cline{1-5} 
 $^{   40}$Ca & $^{  39}$K  &   $1d_{ 3/2}^{-1}$ &     4.33 &         4.51 &
 $^{    }$ &  $^{ 133}$Sn   &   $1h_{ 9/2}$     &   -16.45  &    -13.45\\       \cline{6-10}     
 $^{    }$ &  $^{  39}$K$^{a}$ &   $1d_{ 5/2}^{-1}$ &     5.51 &      5.75 &                 
 $^{ 208}$Pb &  $^{ 207}$Tl &   $2d_{ 3/2}^{-1}$ &    11.33 &    11.93 \\       
 $^{    }$ &  $^{  41}$Sc   &   $1f_{ 7/2}$     &    -8.30 &    -8.78 &                     
 $^{    }$ &  $^{ 207}$Tl$^{a}$ &   $1h_{11/2}^{-1}$ &  25.62  &    27.08 \\   
 $^{    }$ &  $^{  41}$Sc$^{a}$  &   $1f_{ 5/2}$ &   -10.80 &    -10.36 &                    
 $^{    }$ &  $^{ 207}$Tl$^{b}$ &   $2d_{ 5/2}^{-1}$ &    15.59 &16.45 \\ 
 $^{    }$ &  $^{ 39}$Ca    &   $1d_{ 3/2}^{-1}$ &     2.20 &     2.23 &                     
 $^{    }$ & $^{ 209}$Bi    &   $1h_{ 9/2}$ &   -27.71 &    -29.11 \\    
 $^{    }$ & $^{  41}$Ca    &   $1f_{ 7/2}$     &    -3.68 &     -3.66 &                          
 $^{    }$ & $^{ 209}$Bi$^{a}$ &   $2f_{ 7/2}$ &   -21.31 &   -22.28 \\      
 \cline{1-5} 
 $^{  48}$Ca &  $^{  47}$K  &   $1d_{ 3/2}^{-1}$ &     5.90 &     5.97 &
 $^{    }$ &   $^{ 209}$Bi$^{b}$  &   $1i_{13/2}$ & -29.27 &   -30.79\\    
 $^{    }$ &  $^{  47}$K$^{a}$  &   $1d_{ 5/2}^{-1}$ &     7.34 & 7.43 & 
 $^{    }$ &  $^{ 207}$Pb   &   $2f_{ 5/2}^{-1}$ &    11.62 &   11.40 \\  
 $^{    }$ &  $^{  49}$Sc   &   $1f_{ 7/2}$ &   -10.32 &  -10.45 &   
 $^{    }$ & $^{ 207}$Pb$^{a}$  &   $3p_{ 3/2}^{-1}$ &     7.49 &7.37 \\  
 $^{    }$ & $^{ 49}$Sc$^{a}$  &   $1f_{ 5/2}$ &   -10.52 &   -10.63 &   
 $^{    }$ &  $^{ 207}$Pb$^{b}$ &   $1i_{13/2}^{-1}$ &    18.54 &     18.58 \\ 
 $^{    }$ & $^{  47}$Ca    &   $1f_{ 7/2}^{-1}$ &   4.95 &      4.86 &                    
  $^{    }$ &$^{ 207}$Pb$^{c}$   &   $2f_{ 7/2}^{-1}$ &    12.97 &12.80 \\       
 $^{    }$ & $^{  49}$Ca    &   $2p_{ 3/2}$ &  -2.58 &  -2.40 & 
$^{    }$ & $^{ 207}$Pb$^{d}$   &   $1h_{ 9/2}^{-1}$ &    19.53 &  19.80 \\  
\cline{1-5} 
 $^{  60}$Ca & $^{  59}$K   &   $1d_{ 3/2}^{-1}$ &     4.28 &   4.61 &                     
 $^{    }$ & $^{ 209}$Pb   &   $2g_{ 9/2}$ &   -14.42 &   -13.93  \\
 $^{    }$ &  $^{  59}$K$^{a}$  &   $1d_{ 5/2}^{-1}$ &     5.67 &    6.08 & 
 $^{    }$ &  $^{ 207}$Pb   &   $2f_{ 5/2}^{-1}$ &    11.62 &   11.40 \\  
 $^{    }$ &  $^{ 61}$Sc    &   $1f_{ 7/2}$        &    -8.17 &    -8.70 &  
 $^{    }$ &  $^{ 209}$Pb$^{a}$   &   $1i_{11/2}$ &   -21.80 &    -21.58 \\   
 $^{    }$ &  $^{ 61}$Sc$^{a}$  &   $1f_{ 5/2}$ &    -7.63 &  -8.16 & 
 $^{    }$ &  $^{ 209}$Pb$^{b}$  &   $1j_{15/2}$ &   -20.11 &     -19.81  \\ 
 \cline{1-5} \cline{6-10}
 $^{  90}$Zr &  $^{  91}$Nb    &   $1g_{ 9/2}$ &   -15.55 &  -15.94 & 
 & & & & \\
 $^{    }$ &  $^{ 89}$Zr  &   $1g_{ 9/2}^{-1}$ &     8.04 & 8.01 &
 & & & & \\
 $^{    }$ & $^{  91}$Zr   &   $1g_{ 7/2}$ &    -8.62 &    -8.48 & 
 & & & & \\
\hline \hline 
\end {tabular}
}
\end{center}
\vspace*{-0.5cm}
\caption{\small 
Values of $\eqm_{\rm RPA}$, in e$\,$fm$^2$, obtained in our model for the two
interactions considered.  We present the core nuclei, the associated 
$A\pm1$ nuclei and the s.p. states which characterize them. The upper
latin letters indicate the excited states of the isotope. 
}
\label{tab:Q2a}
\end{table}

The role of the continuum in this type of calculations has been
studied in Ref. \cite{don11a} where a detailed discussion about the
position of the centroid energies and the sum rule exhaustion of the
$2^+$ strength distribution is presented, resulting compatible with
the theoretical expectations \cite{lan64}.  There is a difference
between the results of Ref. \cite{don11a} and those of the present
work, since in these latter RPA calculations the spin-orbit and
Coulomb terms of the effective nucleon-nucleon interaction have been
considered by using the procedure of Ref. \cite{don14a}. 
We have verified that for the $\eqm$ values, 
and also for those of $\mdm$, the differences between the two 
types of calculations are of the order of few percents.

If we neglect the cases of the $^{16}$O and $^{40}$Ca nuclei, the
average relative difference between our excitation energies and the
experimental ones is about 15\%.  Even though our RPA description of
the low-lying $2^+$ excitation of the core nuclei is not satisfactory,
we have to consider that, in the calculation of $\eqm$ in the odd-even
nearby nuclei, the relevant quantity to be considered is the global
strength distribution.  Our RPA calculations generate the expected
amount of $2^+$ strength, but it is not properly distributed. This may
affect our $\eqm$ results because the energy denominators of the
kernels (\ref{eq:kern1}) and (\ref{eq:kern2}) indicate that the
low-lying excited states are more important than those with high
energies in the evaluation of the matrix element in
Eq. (\ref{eq:main}). We have tested this sensitivity by artificially
changing of the 10\% the values of the $\omega_\nu$ energies
calculated with the D1M force.  The largest deviations of the $\eqm$
values we have identified are of about 10\%, much smaller in any case
than the differences with the IPM values.

Around the core nuclei, listed in Table \ref{tab:E2}, we have selected
48 states of several isotopes and isotones with one nucleon more or
less, which allow us to compare our results with the experimental data
of the compilation of Ref. \cite{sto05} and also with the results of
Ref. \cite{rin73} obtained within the FFS theory.

In Table \ref{tab:Q2a}, we show the RPA results obtained in our model
by using the D1M and D1S nucleon-nucleon effective interactions. In
this table we indicate the core nuclei, the specific isotopes for
which $\eqm$ was calculated, and the s.p. states characterizing them.
In general, we have considered the states with lowest energy and  
$\eqm \not= 0$. 
For example, the true ground
states of $^{207}$Tl and $^{207}$Pb are a 1/2$^+$ and a 1/2$^-$ states
respectively, but both have $\eqm=0$.  For this reason, in these two
cases, we have indicated 
the 3/2$^+$ and 5/2$^-$ states, respectively. 
In the table, the states of the same isotope with higher 
energy are indicated by a superscript latin letter.

Our calculations confirm the well known fact that the $\eqm$ values of
nuclei with one nucleon more than the doubly closed shell core are
negative, and those with a nucleon less are positive \cite{boh69}. A
first general remark about the results of Table \ref{tab:Q2a}, is that
the values we have obtained by using the two different interactions
are very similar.  The average relative difference between the RPA
results calculated with D1M and D1S interactions is of few percent.
Henceforth, if not explicitly stated, we shall discuss only the
results obtained with the D1M interaction.

\begin{figure}[h]
\begin{center}
\includegraphics[scale=0.4, angle=0] {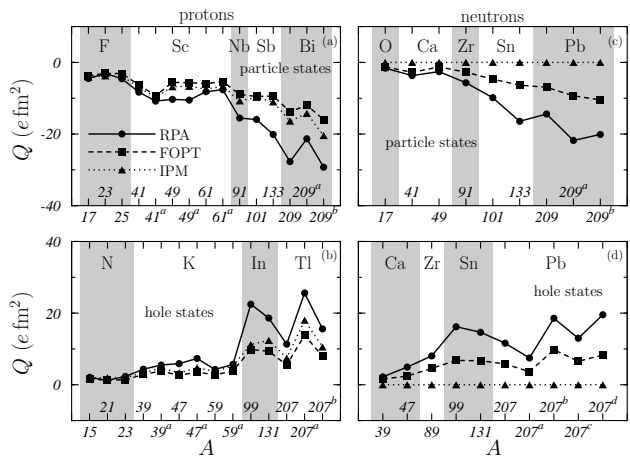} 
\vspace*{-0.3cm}
\caption{\small Values of $\eqm$, in e fm$^2$, for RPA (circles), FOPT
  (squares), and IPM (triangles), calculated with the D1M interaction
  for the various nuclei studied. The left (right) panels correspond
  to proton (neutron) states, while the upper (lower) panels show the
  results for particle (hole) states.  The lines have been drawn to
  guide the eyes.  The latin upper indexes indicate excited states
  following the nomenclature of Table \ref{tab:Q2a}.  The RPA values
  for the nuclei around the $^{116}$Sn core, drawn as open circles,
  have been multiplied by 0.25.  }
\label{fig:rpaper1}
\end{center}
\end{figure}

We compare in Fig. \ref{fig:rpaper1} the results of the IPM,
triangles, FOPT, squares, and RPA, circles, calculations. The four
panels show separately the results obtained for hole and particle-like
states and for protons and neutrons.  Since we do not use effective
charge, the IPM results for the neutron states are zero.

A first, general, remark is that, in absolute value, the RPA results
are always larger than those obtained in the other calculations.  It
is interesting to observe that, in the proton case, the FOPT results
are very close to those of the IPM, and, sometime, even smaller 
in absolute value.  In general, 
the RPA results produce a relatively small correction with
respect to the IPM values.

\begin{figure}[ht]
\begin{center}
\includegraphics[scale=0.4, angle=0] {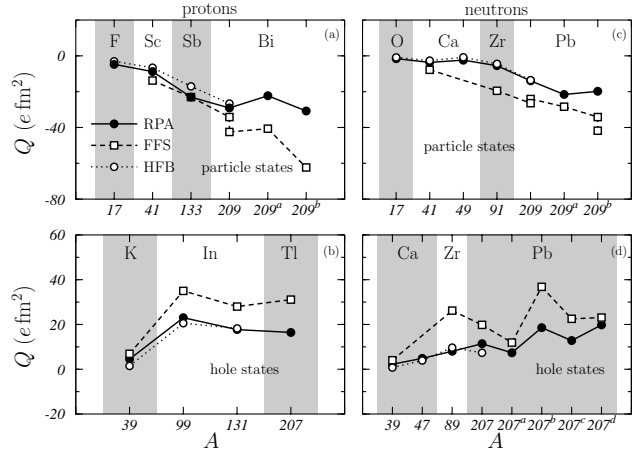} 
\vspace*{-0.3cm}
\caption{
Values of $Q$, in  $e$ fm$^2$, obtained in our RPA calculations with
D1S interaction (solid circles), compared to the FFS results
taken from Refs. \cite{spe77,tol11,tol12,voi12} (open squares)
and to the HFB results found for D1S interaction (open circles).
}
\label{fig:HFB}
\end{center}
\end{figure}

We carried out an additional test of our model by comparing our
results with those of other theoretical approaches, the
Hartree-Fock-Bogolioubov (HFB) and the FFS theory
\cite{spe77,tol11,tol12,voi12}.  This comparison is shown in
Fig. \ref{fig:HFB}.  The HFB calculations have been performed by using
the technique developed by Robledo {\it et al.}~\cite{rob12}, based on
the gradient method. In the minimization procedure both time-even and
time-odd fields are considered in the calculation of the energy
functional and this allows us to avoid the equal filling approximation
\cite{rod11}.  In the figure, we show the RPA and HFB results obtained
with the D1S interaction.  The relevant point for the present study is
the remarkable agreement between the HFB and our RPA results.

The other comparison proposed in Fig. \ref{fig:HFB} is that with the
results of FFS calculations. This last approach is rather similar to
ours, and the differences are due to the use of different
interactions, and s.p. bases.  The results of the nuclei around
$^{208}$Pb taken from Ref. \cite{spe77}, have been obtained by fully
exploiting the philosophy of the Landau-Migdal approach.  The
s.p. wave functions have been generated by using a Woods-Saxon
potential well, and the RPA calculations have been conducted with a
Landau-Migdal interaction whose parameters have been chosen to
reproduce at best the data. The results of nuclei lighter than
$^{208}$Pb, are taken from Refs. \cite{tol11,tol12,voi12}, and have
been obtained by using a slightly different computational
scheme. Skyrme interactions of different type have been used in HF
calculations to generate the s.p. basis wave functions and, in the
second step, also to solve the RPA equations.  The zero-range
characteristics of this type of interaction requires the use of a
cutoff renormalisation parameter whose value depends on the size of
the s.p. configuration space used in the RPA calculations.  The $\eqm$
values obtained with the FFS approach are, in absolute value, larger
than ours.

%
\begin{table}[b]
\begin{center}
\begin{tabular}{rccrrrcc}
\hline \hline 
&  & & & & & \multicolumn{2}{c}{exp} \\ \cline{7-8}
core & nucleus  & s.p. state        &  IPM  &   FOPT &   RPA  & Ref. \cite{sto05} & Ref. \cite{gar15} \\ \hline
$^{16}$O  & $^{17}$F  & $1d_{5/2}$     & -4.12 &  -3.74 &   -4.50 &   -5.80 & \\
         & $^{17}$O  & $1d_{5/2}$     &  0.0 &  -1.51 &   -1.61 &   -2.60 & \\\hline
$^{40}$Ca &$^{39}$K  &  $1d^{-1}_{3/2}$ &  3.37 &   3.04  &    4.33 &    5.90 & \\ 
         & $^{41}$Sc &  $1f_{7/2}$ &   -7.01  &   -6.31 &   -8.30 &  -15.60 & \\
         & $^{39}$Ca &  $1d^{-1}_{3/2}$ & 0.0 &   1.78 &    2.20 &   3.60 &\\ 
         & $^{41}$Ca &  $1f_{7/2}$     & 0.0 &    -3.17 &   -3.68 & -9.00 &\\ \hline
$^{48}$Ca & $^{47}$Ca &  $1f^{-1}_{7/2}$ & 0.0 &    2.43 & 4.95 & 2.10 & {8.46} \\ 
         & $^{49}$Ca &  $2p_{3/2}$ & 0.0 &  -1.24 & -2.58 && {-3.63} \\ \hline 
$^{90}$Zr & $^{91}$Zr & $2d_{5/2}$      & 0.0   &  -3.15  &  -5.68 &  -17.60 & \\ \hline
$^{208}$Pb & $^{209}$Bi & $1h_{9/2}$  &  -16.39 &  -13.84 &  -27.71 &  -51.60& \\
& $^{209}$Pb &  $2g_{9/2}$ &    0.0 &   -7.55 &  -14.42 &  -30.00 & \\
\hline  \hline 
\end {tabular}
\end{center}
\vspace*{-0.5cm}
\caption{\small Values of $\eqm$, in e$\,$fm$^2$, for IPM, FOPT and RPA, 
calculated with the D1M interaction, compared to the experimental data 
taken from the compilation \cite{sto05} and from Ref.~\cite{gar15}.
The s.p. states refer to the corresponding doubly magic cores.
}
\label{tab:exp}
\end{table}

In Table \ref{tab:exp} we compare our results with the experimental 
values taken from the compilation of Ref. \cite{sto05} and with 
those given in Ref. \cite{gar15}.  
The RPA description of these experimental data is not particularly 
satisfactory. All the experimental values are larger than our 
predictions. In heavier nuclei our calculations are able
to account only for about half of the observed $\eqm$ values. This 
indicates that our approach is not fully able to describe the large
collectivity of the $2^+$ excitations. On the other hand, the improvement
with respect to the IPM predictions, and also with respect to the FOPT
predictions, is remarkable.

\subsection{The $\mdm$ and the $1^+$ excitation.}

We use the same set of even-even core nuclei 
 adopted for the calculation of 
$\eqm$ also for the investigation of the magnetic dipole moment
$\mdm$. Around each core nucleus we calculate the $\mdm$ values for
the four nuclei with one nucleon more or less.  We present our
results by following the strategy adopted in the investigation of
$\eqm$. We first discuss the features of the $1^+$ excitation in the
core nuclei and, after, we present the results of the calculation of
$\mdm$ and we compare them with the experimental values.

We present in Table \ref{tab:M1ene} the excitation energies and the
\bmone values of the main low-lying $1^+$ states, and we
compare them with the experimental values taken from
Refs. \cite{led78,bnlw}.  We indicate in the table also the dominant
ph transitions.

\begin{table}[h]
\begin{center}
{\footnotesize
\begin{tabular}{ccccccccccc}
\hline \hline 
&~~~~~~~~~~&  \multicolumn{3}{c}{$\omega$ (MeV)} &~~~~~~~~~~~~~& \multicolumn{2}{c}{\bmone  ($\mu_N^2$)} 
&~~~~~~~~~~& \multicolumn{2}{c}{main s.p. transitions}  \\ \cline{3-5} \cline{7-8} \cline{10-11}
nucleus && D1M &  D1S &  exp && D1M &  D1S && proton & neutron \\ \hline
      $^{16}$O &&17.76 &  18.37 & 13.66  && 0.01 &  0.01  && $(2p_{3/2} 1p^{-1}_{1/2})$ & \\
      $^{22}$O &&  8.35 &   8.97  &            &&  5.81 &  5.18 &&   & $(1d_{3/2} 1d^{-1}_{5/2})$ \\
      $^{24}$O &&  8.50 &  9.06   &  9.50   &&  5.75   & 5.13 &&  & $(1d_{3/2} 1d^{-1}_{5/2})$ \\
    $^{40}$Ca &&14.62 &  15.03 &  9.87   && 0.006  & 0.006  && $(2d_{5/2} 1d^{-1}_{3/2})$ &\\
    $^{48}$Ca &&  9.30 & 10.19  &  10.23 &&   10.23 & 9.66   && & $(1f_{5/2} 1f^{-1}_{7/2})$  \\
    $^{60}$Ca &&  6.73 &  6.76  &             &&  0.04  &  0.04  && & $(3p_{3/2} 2p^{-1}_{1/2})$ \\ 
     $^{90}$Zr &&  9.08 &  9.98  &   9.37   && 13.98 & 13.49  && & $(1g_{7/2} 1g^{-1}_{9/2})$\\
  $^{100}$Sn &&  7.49 &  9.13  &             &&  0.65  &  0.97   && $(1f_{5/2} 1f^{-1}_{7/2})$ & \\
  $^{132}$Sn &&  6.78 &  8.00  &             &&  6.05   &  9.84  & & $(1f_{5/2} 1f^{-1}_{7/2})$ & \\
  $^{208}$Pb &&  6.30 &   7.60 &    5.85  &&  7.80 &  11.89  && $(1i_{11/2} 1i^{-1}_{13/2})$ & $(1h_{9/2} 1h^{-1}_{11/2})$ \\
\hline\hline
\end {tabular}
}
\end{center}
\vspace*{-0.5cm}
\caption{\small 
Energies, $\omega$, and \bmone values
of the main low-lying $1^+$ excited state obtained in our
calculations for the core nuclei considered, with the two interactions we have adopted. 
The experimental values are taken from Refs. \cite{led78,bnlw}.
}
\label{tab:M1ene}
\end{table}

The table shows that the \bmone values of the $^{16}$O, $^{40}$Ca and
$^{60}$Ca nuclei, where all the spin-orbit partner levels are
occupied, are orders of magnitudes smaller than the other ones.  
This confirms the well known fact \cite{boh69,ram91} that the \mone
excitation is strongly excited in nuclei where ph
transitions between spin-orbit partner levels are allowed.  We have presented
in Ref. \cite{co12b} a detailed discussion of the \mone strength
distribution in oxygen and calcium isotopes.

The agreement with the experimental energies is reasonable for almost
all the cases, but for $^{16}$O and $^{40}$Ca. 
We have already pointed out the fact that in the first
two nuclei the $1^+$ excitation is rather weak because all the
spin-orbit partner levels are occupied. In these nuclei it is not easy
to find out the $1^+$ states to be compared with those predicted by
our model, since, experimentally, the \mone strength is rather
fragmented. For example in $^{40}$Ca we found more than twenty $1^+$
states below 12 MeV \cite{led78,bnlw}. 

Our description of the $1^+$ excitation is rather good in the
situations where these excitations are dominated by a few ph
transitions. In these cases our approach works at its best.  The main
features of the transitions are already well described within the IPM
which is corrected by the RPA by including other, less important, ph
transitions.

The ten doubly magic nuclei listed in Table \ref{tab:M1ene} have been
considered as core nuclei where we add or remove one nucleon at the
time to form odd-even nuclei with magnetic dipole moment $\mdm$.
Before discussing the effects of the core polarization, i.e. of the
excitation of the even-even core, we clarify the role played by the
MEC in the calculation of $\mdm$.

\begin{figure}[t]
\begin{center}
\includegraphics[scale=0.4, angle=0] {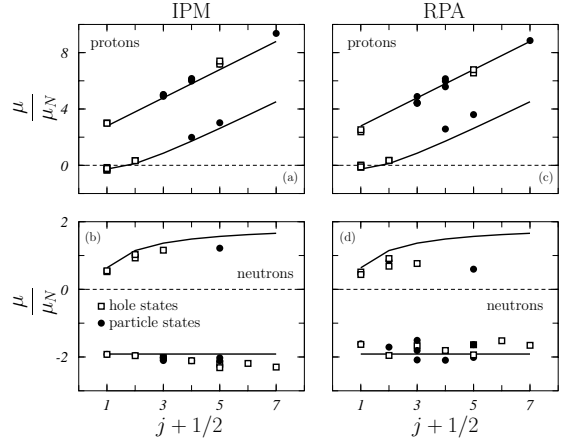}
\vspace*{-0.3cm}
\caption{\small 
Comparison between the values of $\mdm_{\rm IPM}$, 
left panels, and $\mdm_{\rm RPA}$, right panels, 
expressed in nuclear magnetons, and the Schmidt values, as a function of 
the angular momentum $j$ of the odd-even nucleus.  
The results obtained with the full electromagnetic operator, 
$\mop_1=\mop_1^{\rm  OB} + \mop_1^{\rm MEC}$, 
are indicated by the symbols  and the Schmidt values
by the solid lines.
Open squares and solid circles represent the results obtained,
respectively, for 
nuclei characterized by hole or particle-like s.p. states.   
The calculations have been carried out with the D1M interaction.
}
\label{fig:M1MEC}
\end{center}
\end{figure}

In Fig. \ref{fig:M1MEC} we adopt the usual representation of $\mdm$
\cite{rin80} against the angular momentum $j$ of the odd-even nucleus,
to show our results obtained with the D1M force.  In the (a) and (b)
panels the symbols represent the $\mdm_{\rm IPM}$ values obtained by
considering the full electromagnetic operator $\mop_1=\mop_1^{\rm OB}
+ \mop_1^{\rm MEC}$, while the lines indicate the Schmidt values
found for $\mop_1^{\rm OB}$ (see Eq. \eqref{eq:schmidt}).  
The MEC produce relatively small changes
with respect to the Schmidt values, usually of the order of
few percent up to a maximum value of about $10\%$. These changes have a
consistent behavior over all the various regions of the nuclear chart,
and we did not observe remarkable differences between nuclei
characterized by a hole- (open squares) or a particle-like (solid
circles) structure.

\begin{figure}[b]
\begin{center}
\includegraphics[scale=0.4, angle=0] {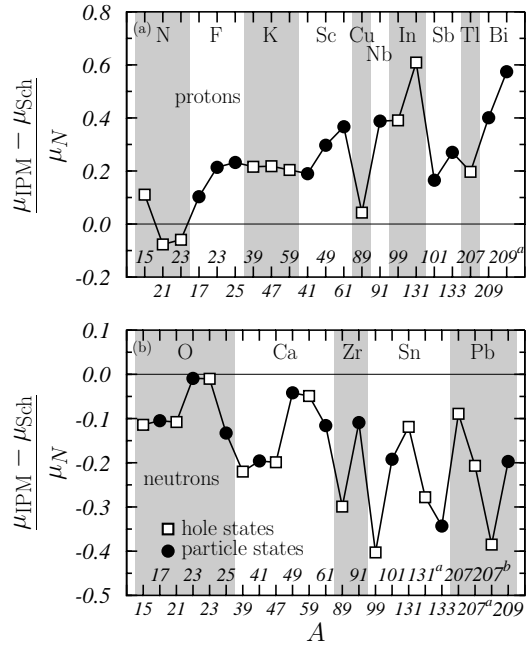}
\vspace*{-0.3cm}
\caption{\small Difference $\mdm_{\rm IPM}-\mdm_{\rm Sch}$, expressed
in nuclear magnetons, for proton, panel (a), and neutron, panel
(b), odd nuclei. 
Open squares and solid circles represent the results obtained 
for nuclei characterized by hole or particle-like structure.
The calculations have been carried out with the D1M interaction.}
\label{fig:M1IPM}
\end{center}
\end{figure}

Another perspective to observe these results is given in
Fig. \ref{fig:M1IPM} where we show the differences $\mdm_{\rm
  IPM}-\mdm_{\rm Sch}$ as a function of the mass number of the nuclei
investigated. The results of panel (a) correspond to the
odd-proton nuclei while those of panel (b) are those obtained for the
odd-neutron nuclei. In the odd-proton cases the MEC always increase
the $\mdm$ values with respect to the Schmidt ones, but for the
$^{21}$N and $^{23}$N nuclei.  The results of panel (b) show that the
effect of the MEC on the odd-neutron nuclei consists in a lowering
with respect to the Schmidt values, without exceptions.
It is possible to outline a general overall tendency of the MEC
effects to increase with the mass number, even though there are quite
a few exceptions.

The effects we have just described are consistent for all the nuclei,
and, in terms of the agreement with the experimental data implies two
contrasting aspects. Since the experimental data are lying between the
two Schmidt lines, the inclusion of the MEC improves the agreement for
odd-proton nuclei with $j_< \equiv l - 1/2$, while they are worsening
it for odd-proton nuclei with $j_> \equiv l+ 1/2$.  We have a similar
situation also for the odd-neutron nuclei where we observe an
improvement in the case of $j_<$ and a worsening for $j_>$.

\begin{table}[h]
\begin{center}
{\scriptsize
\begin{tabular}{ r  llrr r r  llrr r}
\hline \hline 
core & nucleus  &s.p. state & D1M &  D1S & 
& core & nucleus & s.p. state & D1M  & D1S &   \\ 
\hline 
 $^{  16}$O & $^{  15}$N &   1p$_{ 1/2}^{-1}$    &    -0.137  &  -0.142  &    
 & $^{  90}$Zr & $^{  89}$Y &   2p$_{ 1/2}^{-1}$ &    -0.108  &    -0.152 &     \\ 
 $^{  }$ & $^{  17}$F &   1d$_{ 5/2}$       &     4.884  &    4.885 &     
 & $^{  }$ & $^{  91}$Nb &   1g$_{ 9/2}$    &     6.727  &     6.996 & \\ 
 $^{  }$ & $^{  15}$O&   1p$_{ 1/2}^{-1}$       &    0.638   &    0.638  &     
 & $^{  }$ & $^{  89}$Zr &   1g$_{ 9/2}^{-1}$    &    -1.937  &    -1.813  &   \\ 
 $^{  }$ & $^{  17}$O &   1d$_{ 5/2}$       &    -2.088  &    -2.007 &    
 & $^{  }$ & $^{  91}$Zr &   2d$_{ 5/2}$    &    -1.808  &    -1.715 &    \\ \cline{1-6} \cline{7-12}
 $^{  22}$O & $^{  21}$N &   1p$_{ 1/2}^{-1}$    &     0.001  &    -0.042 &
 & $^{ 100}$Sn & $^{  99}$In &   1g$_{ 9/2}^{-1}$ &    6.559  &     6.593  & \\ 
 $^{  }$ & $^{  23}$F &   1d$_{ 5/2}$       &     4.397 &     4.779 &
 & $^{ }$ & $^{ 101}$Sb &   2d$_{ 5/2}$     &     4.448 &     4.476  & \\ 
 $^{  }$ & $^{  21}$O &  1d$_{ 5/2}^{-1}$        &    -1.667  &    -1.487 &
 & $^{ }$ & $^{  99}$Sn &   1g$_{ 9/2}^{-1}$     &    -1.638  &    -1.771  & \\ 
 $^{  }$ & $^{  23}$O &   2s$_{ 1/2}$       &    -1.610  &    -1.485 &
 & $^{ }$ & $^{ 101}$Sn &   2d$_{ 5/2}$     &    -1.507 &     -1.614  & \\ \cline{1-6} \cline{7-12}
 $^{  24}$O & $^{  23}$N &   1p$_{ 1/2}^{-1}$    &      0.013&     -0.024  &
& $^{ 132}$Sn & $^{ 131}$In &   1g$_{ 9/2}^{-1}$ &     6.788 &    6.767 & \\ 
  $^{  }$ & $^{  25}$F &   1d$_{ 5/2}$       &     4.444  &     4.801 &
 & $^{ }$ & $^{ 133}$Sb &   1g$_{ 7/2}$     &     2.579  &     2.546 &      \\
  $^{  }$ & $^{  23}$O &   2s$_{ 1/2}^{-1}$       &    -1.628  &    -1.482 &
& $^{ }$ & $^{ 131}$Sn &   2d$_{ 3/2}^{-1}$     &     0.684  &     0.758  &     \\ 
  $^{  }$ & $^{  25}$O &  1d$_{ 3/2}$        &     0.842  &     0.841 &
 & $^{ }$ & $^{ 131}$Sn$^a$ &   1h$_{11/2}^{-1}$ &    -1.520  &    -1.688  &   \\ 
 \cline{1-6} 
 $^{  40}$Ca & $^{  39}$K &   1d$_{ 3/2}^{-1}$    &     0.358  &     0.350 &   
 & $^{ }$ & $^{ 133}$Sn &   1h$_{ 9/2}$     &     0.596  &     0.746 & \\ \cline{7-12}
 $^{  }$ & $^{  41}$Sc &   1f$_{ 7/2}$      &     5.971  &     5.972 &   
& $^{ 208}$Pb & $^{ 207}$Tl &   3s$_{ 1/2}^{-1}$ &     2.528  &     2.538 &   \\ 
  $^{  }$ & $^{  39}$Ca &   1d$_{ 3/2}^{-1}$      &     0.910  &     0.918 &   
 & $^{ }$ & $^{ 209}$Bi &   1h$_{ 9/2}$     &     3.604  &     3.548 &    \\ 
  $^{  }$ & $^{  41}$Ca &   1f$_{ 7/2}$      &    -2.096  &    -2.098  &   
 & $^{ }$ & $^{ 209}$Bi$^a$ &   1i$_{13/2}$ &     8.859 &  \\ 
 \cline{1-6} 
 $^{  48}$Ca & $^{  47}$K &   2s$_{ 1/2}^{-1}$    &     2.384  &     2.695 &     
 & $^{ }$ & $^{ 207}$Pb &   3p$_{ 1/2}^{-1}$     &     0.437  & 0.463 &    \\ 
 $^{  }$ & $^{  49}$Sc &   1f$_{ 7/2}$      &     5.577 &    5.888   &
 & $^{ }$ & $^{ 207}$Pb$^a$ &   2f$_{ 5/2}^{-1}$ &     0.764  &     0.857 &   \\
  $^{  }$ & $^{  47}$Ca &   1f$_{ 7/2}^{-1}$      &    -1.809  &    -1.664  &    
 & $^{ }$ & $^{ 207}$Pb$^b$ &   1i$_{13/2}^{-1}$ &    -1.655 &    \\ 
 $^{  }$ & $^{  49}$Ca &   2p$_{ 3/2}$      &    -1.709 &    -1.605 &
 & $^{ }$ & $^{ 209}$Pb &   2g$_{ 9/2}$     &    -1.635 &    -1.756 &  \\ 
 \cline{1-6}  \cline{7-12}
 $^{  60}$Ca & $^{  59}$K &   1d$_{ 3/2}^{-1}$   &     0.347  &  0.338 &
 & & & & & \\
  $^{  }$ & $^{  61}$Sc &  1f$_{ 7/2}$       &     6.146 &        6.147 & 
 & & & & & \\
 $^{  }$ & $^{  59}$Ca &   2p$_{ 3/2}^{-1}$      &    -1.954  &    -1.951 &
 & & & & & \\
 $^{  }$ & $^{  61}$Ca &   1g$_{ 9/2}$      &    -2.015 &      -2.011 &
 & & & & & \\
\hline\hline
\end {tabular}
}
\end{center}
\vspace*{-0.5cm}
\caption{\small 
Values of $\mdm$, in nuclear magnetons, obtained in our RPA model 
with the full $\mop_1^{\rm \, OB} + \mop_1^{\rm \, MEC}$ operator
for the two interactions considered.
We present the core nuclei, the specific $A\pm1$ nuclei 
and the s.p. states which characterise them. The upper 
latin letters indicate the excited states of the isotope. 
}
\label{tab:M1}
\end{table}

In Table \ref{tab:M1} we present the RPA results obtained with the two
interactions we have adopted, and by using the full electromagnetic
operator which includes the MEC.  We have considered those
configurations which generate the ground state of the odd-even nuclei,
and some excited state of interest. In the table, our results are
compared to the available data taken from the compilation of
Ref.~\cite{sto05}.

With the exceptions of the $^{21}$N, $^{23}$N, and $^{131}$Sn cases,
some of which we shall discuss in more detail, the relative
differences between the D1M and D1S results are of the order of
1\%. This indicates that the effects we are discussing are almost
independent of the specific implementation of the interaction used.

The quantitative stability of our results can be observed by
considering that, in our approach, the $^{23}$O nucleus can be
obtained by adding a neutron to the $^{22}$O core or subtracting one
to $^{24}$O. The differences between the two procedures are related to
the different s.p. states produced by the HF calculations and the
results obtained for D1M and D1S interactions show $\mu$ values that
differ on the third and fourth significant figures, respectively.

We compare the D1M results of Table \ref{tab:M1} with the Schmidt
values in the panels (c) and (d) of Fig. \ref{fig:M1MEC}.  In general,
the $\mdm_{\rm RPA}$ values are situated between the two Schmidt
lines, as it is observed for the experimental values. The deviation
from the Schmidt values is particularly evident in the neutron case
for the nuclei with $j_<$ above 5/2, which correspond to the
$^{207}$Pb$^a$ and $^{133}$Sn isotopes. 

\begin{figure}[ht]
\begin{center}
\includegraphics[scale=0.4, angle=0] {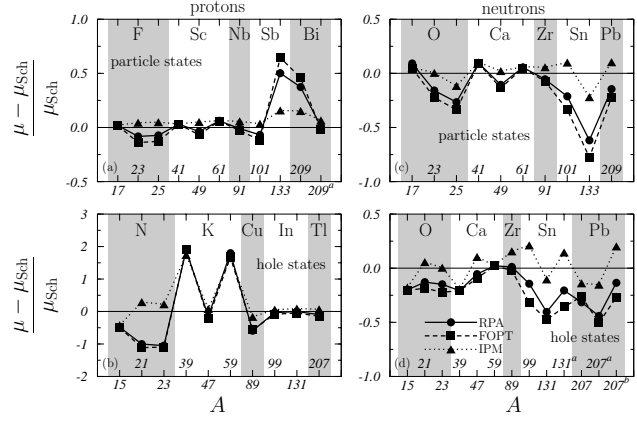} 
\vspace*{-0.3cm}
\caption{\small 
Comparison between our RPA, circles, FOPT,
squares, and IPM, triangles, results. 
The calculations have been carried out with the
D1M interaction and by using the full electromagnetic operator 
$\mop_1=\mop_1^{\rm OB}  + \mop_1^{\rm MEC}$. 
The lines have been drawn to guide the eyes.
}
\label{fig:M1RPA}
\end{center}
\end{figure}

In Fig. \ref{fig:M1RPA} we compare the RPA, FOPT and IPM
results. All the calculations have been carried out with the D1M 
interaction, and by using the complete electromagnetic operator 
$\mop_1=\mop_1^{\rm OB}  + \mop_1^{\rm MEC}$. The results
are shown as relative difference with respect to the Schmidt 
values, Eq. \eqref{eq:schmidt}.
We observe the great similarity between our RPA results and those
obtained in FOPT. This indicates that the structure of the $1^+$ states
is not as collective as that of the $2^+$ states, but it is rather dominated 
by single-particle excitations. 

The FOPT and RPA results show very similar behaviors. In general, the
FOPT deviate more from the Schmidt values than the RPA ones. We remark
that the FOPT approach, often utilized in the literature
\cite{bli53,ari54,ari11,li11,li13}, performs much better in this case
than for the electric quadrupole moment case, as it is shown in
Fig. \ref{fig:HFB}.

Our results agree with those of the non-relativistic calculations of
Refs. \cite{li13,ari11}. More precisely, and always with respect to
the Schmidt values, we observe that: (i) both MEC and RPA increase the
$\mdm$ values in $^{209}$Bi; (ii) in $^{207}$Tl we find an enhancement
due to the MEC while the RPA effects produce a lowering of $\mdm$;
(iii) in $^{209}$Pb we have opposite effects, and (iv) in $^{207}$Pb
both MEC and RPA lower the $\mdm$ values.

The $\mu$ values obtained in RPA show a remarkable difference
with respect to the IPM results. The case of the $^{39}$K and $^{59}$K
nuclei
is particularly interesting since the three calculations provide very similar
results and all of them noticeably deviate from the Schmidt values. This
is only due to the presence of MEC in the operator. This effect should
be investigated better, by considering also that the $^{47}$K nucleus 
does not present this feature. 

\begin{table}[b]
\begin{center}
{\scriptsize
\begin{tabular}{ r  llrrrrrr}
\hline \hline 
core & nucleus  &s.p. state & Schmidt & IPM  & FOPT & RPA & FFS & exp  \\
\hline 
 $^{   16}$O   &   $^{  15}$N & 1p$_{ 1/2}^{-1}$  & -0.264 & -0.153 &  -0.133 &  -0.137 & ---\phantom{--} &  -0.283 \\
                     &   $^{  17}$F  & 1d$_{ 5/2}$          &  4.793 & 4.896 &   4.882 &   4.884 & ---\phantom{--} &   4.721 \\
                     &   $^{  15}$O & 1p$_{ 1/2}^{-1}$   &  0.638 & 0.524 &   0.506 &   0.510 & ---\phantom{--} &   0.720 \\
                     &   $^{  17}$O & 1d$_{ 5/2}$          & -1.913 & -2.018 &  -2.003 &  -2.088 & ---\phantom{--} &  -1.894 \\
\hline
 $^{   40}$Ca &   $^{  39}$K  & 1d$_{ 3/2}^{-1}$  &   0.124 &  0.340 &   0.361 &   0.358 &   0.329 &   0.391 \\
                     &   $^{  41}$Sc & 1f$_{ 7/2}$           &   5.793 &  5.983 &   5.963 &   5.971 &   5.485 &   5.431 \\
                     &   $^{  39}$Ca & 1d$_{ 3/2}^{-1}$  &   1.148 &   0.928 &   0.910 &   0.910 &   0.888 &   1.022 \\
                     &   $^{  41}$Ca & 1f$_{ 7/2}$          &  -1.913 & -2.109 &  -2.088 &  -2.096 &  -1.626 &  -1.595 \\
\hline
 $^{  48}$Ca  &   $^{  47}$K &  2s$_{ 1/2}^{-1}$  &   2.793 & 3.011 &   2.173 &   2.384 & ---\phantom{--} &   1.933 \\
                     &   $^{  47}$Ca & 1f$_{ 7/2}^{-1}$  &  -1.913  & -2.112 &  -1.732 &  -1.809 &  -1.556 &  -1.380 \\
                     &   $^{  49}$Ca & 2p$_{ 3/2}$         &  -1.913 & -1.954 &  -1.654 &  -1.709 &  -1.156 & ---\phantom{--} \\
\hline
$^{  90}$Zr    &   $^{  89}$Y   & 2p$_{ 1/2}^{-1}$   &  -0.264 &  -0.221 &  -0.114 &  -0.108 &  -0.125 &  -0.137 \\
                     &   $^{  89}$Zr & 1g$_{ 9/2}^{-1}$    &  -1.913 & -2.212 &  -1.872 &  -1.937 &  -1.304 &  -1.046 \\
                     &   $^{  91}$Zr & 2d$_{ 5/2}$           &  -1.913 & -2.022 &  -1.769 &  -1.808 &  -1.214 &  -1.304 \\
\hline
 $^{ 132}$Sn &   $^{ 133}$Sb & 1g$_{ 7/2}$         &   1.717 &  1.987 &   2.826 &   2.579 &   2.693 &   3.000 \\
                     &   $^{ 131}$Sn & 2d$_{ 3/2}^{-1}$  &   1.148 &  1.029 &   0.601 &   0.684 &   0.681 &   0.747 \\
             &   $^{ 131}$Sn$^a$ & 1h$_{11/2}^{-1}$ &  -1.913 & -2.191 &  -1.227 &  -1.520 & ---\phantom{--} &  -1.276 \\
\hline
 $^{ 208}$Pb &   $^{ 207}$Tl & 3s$_{ 1/2}^{-1}$   &   2.793 &  2.990 &   2.379 &   2.528 &   1.857 &   1.876 \\
                     &   $^{ 209}$Bi & 1h$_{ 9/2}$          &   2.624 &  3.025 &   3.827 &   3.604 &   3.691 &   4.110 \\
                     &   $^{ 207}$Pb & 3p$_{ 1/2}^{-1}$  &   0.638 &  0.549 &   0.472 &   0.437 &   0.473 &   0.593 \\
               &   $^{ 207}$Pb$^a$ & 2f$_{ 5/2}^{-1}$ &   1.366 &  1.159 &   0.677 &   0.764 &   0.720 &   0.800 \\
                     &   $^{ 209}$Pb & 2g$_{ 9/2}$         &  -1.913 & -2.110 &  -1.486 &  -1.635 &  -1.337 &  -1.474 \\
\hline \hline
\end {tabular}
}
\end{center}
\vspace*{-0.5cm}
\caption{\small 
Comparison between $\mdm$ values, expressed in nuclear magnetons,
calculated in IPM, FOPT and RPA by using the complete electromagnetic
operator  $\mop_1^{\rm \, OB} + \mop_1^{\rm \, MEC}$ with 
the Schmidt values \eqref{eq:schmidt},
the FFS results of Ref. \cite{bor08} and the experimental data
of Ref. \cite{sto05}. 
}
\label{tab:M1exp}
\end{table}
A direct comparison between our results and the available experimental
data taken from Ref.~\cite{sto05} is presented in Table \ref{tab:M1exp}.
In this table we also include the results of the FFS calculation of 
Ref. \cite{bor08}. 
We observe that the effects of the core polarisation become
more important in the heavier nuclei. The differences between
the results of IPM, FOPT and RPA calculations are rather
small in the nuclei around $^{16}$O and $^{40}$Ca.

In the great majority of the cases the RPA results are closer to the
experimental values than the other ones, indicating that the inclusion
of the core polarization improves the agreement with the available
experimental data.  However, this is not a universal behavior since
there are very specific cases, where the
IPM approach produce a better description of the observed $\mdm$
values.  As expected the FFS results, which use effective
electromagnetic operators, are in general closer to the experimental
values that ours.

\begin{table}[t]
\begin{center}
\begin{tabular}{c c c c c c}
\hline \hline
nucleus & interaction & Schmidt & IPM & FOPT & RPA \\
\hline
 $^{21}$N &D1M & -0.264 & -0.341 & 0.030 &  0.001 \\
&D1S & -0.264 & -0.339 & 0.003 & -0.042 \\
\hline 
 $^{23}$N&D1M & -0.264 & -0.323 &  0.025 &  0.013 \\
&D1S & -0.264 & -0.322 & -0.002 & -0.024 \\
\hline \hline
\end{tabular}
\end{center}
\vspace*{-0.5cm}
\caption{\small Values of $\mdm$, in nuclear magnetons, of the
  $^{21}$N and $^{23}$N nuclei. The Schmidt values are compared to
  those found within IPM, FOPT and RPA approaches with the full
  $\mop_1^{\rm \, OB} + \mop_1^{\rm \, MEC}$ operator.  }
\label{tab:M1tens}
\end{table}

We conclude this section by analyzing in detail the cases of the
$^{21}$N and $^{23}$N nuclei. In Table \ref{tab:M1tens} we show the
$\mdm$ values for these two nuclei calculated by
using different approximations.  We observe that the Schmidt values
have negative sign.  The inclusion of the MEC in the IPM calculations
gives a negative contribution, further lowering these values.  The
core polarization plays an important role as the results of two last
columns indicate. In both nuclei, the effect of the residual
interaction has opposite sign with respect to the contribution of the
MEC. This effect is larger in the FOPT than in RPA results. In RPA
calculations, the effects of the D1M force are large enough to change
the sign of $\mdm$ with respect to the Schmidt value. 
Experimental $\mdm$ values for these two nuclei are not available.  In
any case, our RPA results follow the trend of the know experimental
values of lying between the Schmidt values.

\section{Conclusions}
\label{sec:conc}
In this work, we have presented the results of a parameter free
theoretical approach which describes the properties of odd-even nuclei
with one nucleon more, or less, than a doubly magic one.  The
convergence of the energy sum in Eq. (\ref{eq:main}), the fundamental
equation of our model, is ensured by the use of finite-range
nucleon-nucleon effective interaction.  The use of zero-range
interactions, as it is done for example in
Refs. \cite{rin73,bau73,spe77}, requires the inclusion of additional
free parameters.  The universality of our approach allows us to apply
it in each region of the nuclear chart, from light to heavy nuclei.

We have considered two parameterizations of the finite-range effective
Gogny nucleon-nucleon interaction taken from the literature, the D1S
and D1M.  Our calculations are fully self-consistent, meaning that
all results have been obtained by using the same interaction in each
of the three steps of the calculation: the generation of the
s.p. configuration space by means of a HF procedure, the solution of
the RPA equations and, finally, the evaluation of the response of the
odd-even nucleus to the external probe. In each step of the 
calculation the complete interaction has been used, including
the spin-orbit and the Coulomb terms, commonly neglected in RPA. 

We have tested the validity of our approach by calculating the values
of the electric quadrupole moment, $\eqm$, of 48 different states, and
the magnetic dipole moment, $\mdm$, of  44 states, in various
regions of the nuclear chart.  We have used the traditional one-body
operator for the calculation of $\eqm$, while for the $\mdm$ we have
considered also the contribution of MEC.  We use bare electromagnetic
operators whose coupling constants are those of free nucleons.

In absolute value, the results obtained by considering the MEC are
always larger than the Schmidt values, implying an improvement of the
agreement with experimental data for odd-proton nuclei with $j_< $ and
with odd-neutron nuclei with $j_>$, and a worsening for the other type
of nuclei.

The differences between the results obtained with the two forces are
small, negligible if compared to the differences between RPA
and IPM results. This indicates that the quantitative
description of the core polarisation effects in our calculations is
more related to the theoretical approach, rather than to the use of a
specific interaction.

Our calculations require the description of the excitation of $2^+$
and $1^+$ states in doubly closed shell nuclei, in the framework of
the RPA theory. We observe that our 
model describe better the  $1^+$ excitations than the $2^+$ ones. 
This occurs because the most important $1^+$ states, those
where the largest part of the strength is concentrated, are dominated
by a single ph transition. 
In this case, our approach works at its best, by correcting the 
main ph transition with the presence of other one-particle one-hole
transitions. On the other hand,
the structure of many of the $2^+$ states is very collective, and 
a good description of it requires to go beyond the linear combination 
of one-particle one-hole transitions which is what the RPA 
considers.

This difference in the description of the two multipolarities is
reflected in the results we have obtained for the values of $\eqm$ and
$\mdm$. The description of the magnetic dipole moments is certainly
better than that of the electric quadrupole moments. 
The similarity between our results and those obtained in FOPT
is a further indication of the perturbative structure of the $1^+$ 
excitations.  In any case, we have to point
out that our results show a remarkable improvement of the description
of the experimental values with respect to the IPM predictions, 
and also with respect to the FOPT for $\eqm$.

We found a good agreement between our results and those of the FFS
theory given in the literature
\cite{rin73,bau73,spe77,bor08,bor10,tol11,tol12,voi12}. The basic
theoretical hypotheses of the two approaches are rather similar.  More
remarkable is the agreement with the $\eqm$ values obtained 
in the HFB calculations.  
Our approach and the HFB treat the core polarisation in a
completely different manner, but they produce similar results.
In our approach the IPM response is corrected by the  
core polarisation described in terms of particle-hole 
excitations. Alternatively the HFB theory corrects the IPM
by considering the effects of the pairing  force. These two 
alternative visions of effects beyond the IPM picture generate
very similar quantitative results. It would be interesting to
identify specific observables which allow us to disentangle
the two effects.

Despite the limitations we have pointed out, we think that
our parameter free, self-consistent approach is adequate to make
predictions of the properties of odd-even nuclei nearby the doubly
magic-ones in regions of the nuclear chart not yet experimentally
explored.

\appendix
\section{MEC matrix elements}
\label{sec:mec}

In this appendix we give the explicit expressions of the matrix
elements of the MEC inserted in Eq. (\ref{eq:main}) to calculate 
the values of $\mdm$. We have considered the contribution of 
the two diagrams indicated in Fig.~\ref{fig:MEC} and called them
seagull and pionic. We give here the final results of a calculation which describes
these contributions in terms of Feynman's diagrams, makes a
non-relativistic reduction, a multipole expansion and the use of
s.p. wave functions with the angular coupling indicated in Eq. 
\eqref{eq:spwf}.

We call
$q$ the modulus of the momentum that the photon is exchanging with the
nucleus, and $\omega$ the photon energy.
In the specific case under study we have $\omega=\epsilon_{\alpha \beta}$ and
$q=\omega / (\hbar c)$.

The contribution of the seagull term is given by
\begin{eqnarray}
\nonumber
\langle  \,\phi_{a}\, \| \,\mop^{\rm s}_J \, \|\,
\phi_{b} \, \rangle & = &
\xi(l_\alpha + l_\beta +J+1) \, 
(-1)^{j_a - 1/2}
\, \frac{\what{j_a} \what{j_b} \what{J}} {\sqrt{J(J+1)}}
\, \threej {j_a} {j_b} {J} {\half} {-\half} {0}
\\
\nonumber && \times
\left[G^P_E(q,\omega) \,-\, G^N_E(q,\omega) \right] \,
\int {\rm d}r\, r^2 \, j_J(qr) \,{\cal F}^{(\rm s)}_{a,b,J}(r) \, ,
\end{eqnarray}
where $G^P_E$ and $G^N_E$ are, respectively, the proton and neutron 
electromagnetic form factors, and we have defined
\begin{eqnarray}
{\cal F}^{(\rm s)}_{a,b,J}(r) &=&
- 2\, \frac{f^2_\pi}{m^2_\pi} \,
\sum_{h<\epsilon_{\rm F}} \, \what{j}_h^2 \,
\left[\delta_{ab,p} \delta_{h,n} - \delta_{ab,n} \delta_{h,p} \right]
\\ \nonumber
&& \times 
\left\{ \sum_{L_1} \xi(l_a + l_h + L_1 +1 ) \, \what{L}^2_1 \, 
\threej {j_a} {j_h} {L_1} {\half} {-\half} {0}^2\, 
\kappa_{a b}\,  {\cal T}_{a h}(L_1,r) \, R_h(r) \, R_b(r) \right.
\\ 
&& \,\,\,\,\,\,\,\,\,\,\,\, + \, 
\nonumber 
\left.
\sum_{L_2} \, \xi(l_b + l_h + L_2 +1 )\, \what{L}^2_2 \,
\threej {j_h} {j_b} {L_2} {\half} {-\half} {0}^2 \,
\kappa_{a b} \, {\cal T}_{h b}(L_2,r) \,R_a(r) \,R_h(r)
\right\} \, ,
\end{eqnarray}
where $f^2_\pi=0.079$ is the pion-nucleon coupling constant, 
$m_\pi$ is the pion mass,  $\kappa$ is defined in Eq. \eqref{eq:kappa},
$\xi(l)=1$ if $l$ is even and zero otherwise, $R(r)$ is the radial
part of the s.p. wave function \eqref{eq:spwf}, and ${\cal T}$ is
defined as
\begin{eqnarray}
\nonumber
{\cal T}_{a b}(L,x) &=&
\int_0^\infty {\rm d}r \, r^2 \, 
\left\{\left(\frac{\rm d}{{\rm d}r} \,+\, \frac{\kappa_{a b} +2}{r} \right)\,
 R_a(r)\, R_b(r) \right\} 
\\ 
&& \times
\int_0^\infty {\rm d}k \, k^2 \,\frac {2}{\pi} \, v_\pi(k,\epsilon_{a b}) \,
j_L(kr)\, j_L(kx) \, .
\label{eq:calt}
\end{eqnarray}

In the above equations we have indicated with $v_\pi$ the pion
propagator
\beq
v_\pi(k,\epsilon) \,= \,
\frac {1}{k^2 - \epsilon^2 + m^2_\pi} \, .
\eeq

The contribution of the pionic term is 
\beq
\langle  \,\phi_{a}\, \| \,\mop^{\pi}_J \, \|\,
\phi_{b} \, \rangle \,
= \,
\xi(l_\alpha + l_\beta +J+1) \, 
\what{j_a} \, \what{j_b} \, \what{J} \, F_{\pi \gamma} (q,\omega)\, 
\int {\rm d}r\, r^2 \,  j_J(qr) \,{\cal F}^{(\pi)}_{a,b,J}(r) \, ,
\eeq
where 
\beq
F_{\pi \gamma} (q,\omega) \, = \, \frac{1}{1+(q^2 - \omega^2)/m_\rho^2} \, 
\eeq
is the electromagnetic form factor
of the pion, with $m_\rho=776.0\,$MeV the $\rho$-meson mass, and 
\begin{eqnarray}
{\cal F}^{(\pi)}_{a,b,J}(r) &=&
4\, \frac{f^2_\pi}{m^2_\pi} \,\frac{1}{r}\,
\sum_{h<\epsilon_{\rm F}} \,(-1)^{j_h+j_b} \, \what{j}^2_h \, 
\left[\delta_{ab,p} \delta_{h,n} - \delta_{ab,n} \delta_{h,p} \right]
\\ \nonumber && \times
\sum_{L_1 L_2} \,\xi(l_a + l_h + L_1 +1 ) \,\xi(l_b + l_h + L_2 +1 )  \,
\what{L}^2_1 \,\what{L}^2_2 
\\ \nonumber && \times
\sixj{L_1}{L_2}{J}{j_b}{j_a}{j_h}\,
\threej{j_a}{j_h}{L_1}{\half}{-\half}{0}\,
\threej{j_h}{j_b}{L_2}{\half}{-\half}{0}\,
\threej{L_1}{J}{L_2}{1}{-1}{0}\,
\\ \nonumber && \times
\sqrt{L_2(L_2+1)}\, {\cal T}_{a h}(L_1,r) \, {\cal T}_{h b}(L_1,r) \, .
\end{eqnarray}

\acknowledgments
This work has been partially supported by the Junta de Andaluc\'{\i}a
(FQM0220), the European Regional Development Fund (ERDF) and the
Spanish Ministerio de Econom\'{\i}a y Competitividad (FPA2012-31993).


\begin{thebibliography}{10}
\expandafter\ifx\csname url\endcsname\relax
  \def\url#1{\texttt{#1}}\fi
\expandafter\ifx\csname urlprefix\endcsname\relax\def\urlprefix{URL }\fi

\bibitem{goo72}
P.~Goode, B.~J. West, S.~Siegel, Nucl. Phys. A 187 (1972) 249.

\bibitem{ell77}
P.~J. Ellis, E.~Osnes, Rev. Mod. Phys. 49 (1977) 777.

\bibitem{li13}
J.~Li, J.~X. Wei, J.~N. Hu, P.~Ring, J.~Meng, Phys. \ Rev. \ C 88 (2013)
  064307.

\bibitem{mig67}
A.~Migdal, Theory of finite Fermi systems and applications to atomic nuclei,
  Interscience, London, 1967.

\bibitem{rin73}
P.~Ring, R.~Bauer, J.~Speth, Nucl. \ Phys. A 206 (1973) 97.

\bibitem{bau73}
R.~Bauer, J.~Speth, V.~Klemt, E.~Werner, T.~Yamazaki, Nucl. \ Phys. A 209
  (1973) 535.

\bibitem{spe77}
J.~Speth, E.~Werner, W.~Wild, Phys. \ Rep. 33 (1977) 127.

\bibitem{bor08}
I.~N. Borzov, E.~E. Saperstein, S.~V. Tolokonnikov, Physcs \ of \ Atomic \
  Nuclei 71 (2008) 469.

\bibitem{bor10}
I.~N. Borzov, E.~E. Saperstein, S.~V. Tolokonnikov, G.~Neyens, N.~Severijns,
  Eur. \ Phys. \ J. \ A 45 (2010) 159.

\bibitem{tol11}
S.~V. Tolokonnikov, S.~Kamerdzhiev, D.~Voitenkov, S.~Krewald, E.~E. Saperstein,
  Phys. \ Rev. \ C 84 (2011) 064324.

\bibitem{tol12}
S.~V. Tolokonnikov, S.~Kamerdzhiev, S.~Krewald, E.~E. Saperstein, D.~Voitenkov,
  Eur. \ Phys. \ J. \ A 48 (2011) 70.

\bibitem{voi12}
D.~Voitenkov, O.~Achakovskiy, S.~Kamerdzhiev, S.~V. Tolokonnikov,
  arxiv:1210.1691 [nucl-th].

\bibitem{cha07t}
F.~Chappert, Nouvelles param\'etrisation de l'interaction nucl\'eaire effective
  de {Gogny}, Ph.D. thesis, Universit\'e de Paris-Sud XI (France),
  http://tel.archives-ouvertes.fr/tel-001777379/en/ (2007).

\bibitem{don14a}
V.~{De Donno}, G.~Co', M.~Anguiano, A.~M. Lallena, Phys. \ Rev. \ C 89 (2014)
  014309.

\bibitem{fet71}
A.~L. Fetter, J.~D. Walecka, Quantum theory of many-particle systems,
  McGraw-Hill, S. Francisco, 1971.

\bibitem{rin80}
P.~Ring, P.~Schuck, The nuclear many-body problem, Springer, Berlin, 1980.

\bibitem{edm57}
A.~R. Edmonds, Angular momentum in quantum mechanics, Princeton University
  Press, Princeton, 1957.

\bibitem{sch37}
T.~Schmidt, Z. Phys. 106 (1937) 358.

\bibitem{boh69}
A.~Bohr, B.~R. Mottelson, Nuclear structure, vol. I, Benjamin, New York, 1969.

\bibitem{ber91}
J.~F. Berger, M.~Girod, D.~Gogny, Comp. \ Phys. \ Commun. 63 (1991) 365.

\bibitem{gor09}
S.~Goriely, S.~Hilaire, M.~Girod, S.~P\'eru, Phys. \ Rev. \ Lett. 102 (2009)
  242501.

\bibitem{don14b}
V.~{De Donno}, G.~Co', M.~Anguiano, A.~M. Lallena, Phys. \ Rev. \ C 90 (2014)
  024326.

\bibitem{del10}
J.-P. Delaroche, M.~Girod, J.~Libert, H.~Goutte, S.~Hilaire, S.~P\'eru,
  N.~Pillet, G.~F. Bertsch, Phys. \ Rev. \ C 81 (2010) 014303.

\bibitem{ang14}
M.~Anguiano, A.~M. Lallena, G.~Co', V.~{De Donno}, J. \ Phys. \ G 41 (2014)
  025102.

\bibitem{led78}
C.~M. Lederer, V.~S. Shirley, Table of isotopes, 7th ed., John Wiley and sons,
  New York, 1978.

\bibitem{bnlw}
{Brookhaven National Laboratory}, National nuclear data center,
  http://www.nndc.bnl.gov/.

\bibitem{ram01}
S.~Raman, C.~W. {Nestor Jr.}, P.~Tikkanen, Phys. \ Rev. \ C 41 (1990) 1084.

\bibitem{don11a}
V.~{De Donno}, G.~Co', M.~Anguiano, A.~M. Lallena, Phys. \ Rev. \ C 83 (2011)
  044324.

\bibitem{lan64}
A.~M. Lane, Nuclear theory, W. A. Benjamin, New York, 1964.

\bibitem{sto05}
N.~J. Stone, Atomic \ Data \ and \ Nuclear \ Data \ Tables 90 (2005) 75.

\bibitem{rob12}
L.~M. Robledo, R.~Bernard, G.~F.~Bertsch, Phys. \ Rev. \ C 86 (2012) 064313.

\bibitem{rod11}
R.~Rodr\'{\i}guez-Guzm\'an, P.~Sarriguren, L.~M. Robledo, Phys. \ Rev. \ C 83
  (2011) 044307.

\bibitem{gar15}
R.~F. {Garcia Ruiz}, et~al., arxiv:1504.04474 [nucl-exp].

\bibitem{ram91}
S.~Raman, L.~W. Fagg, R.~S. Hicks, Giant magnetic resonances, in Electric and
  magnetic giant resonances in nuclei, {\rm J. Speth ed.}, World Scientific,
  Singapore, 1991.

\bibitem{co12b}
G.~Co', V.~{De Donno}, M.~Anguiano, A.~M. Lallena, Phys. \ Rev. \ C 85 (2012)
  034323.

\bibitem{bli53}
R.~J. Blin-Stoyle, Proc. Phys. Soc. A 66 (1953) 1158.

\bibitem{ari54}
A.~Arima, H.~Horie, Prog. Part. Nucl. Phys. 12 (1954) 623.

\bibitem{ari11}
A.~Arima, Sci. China Phys. Mec. Astron. 54 (2011) 188.

\bibitem{li11}
J.~Li, J.~M. Yao, J.~Meng, A.~Arima, Prog. Theor. Phys. 125 (2011) 1185.

\end{thebibliography}
%

%
\end{document}